\documentclass[runningheads]{llncs}
\newif\ifarxiv
\arxivtrue  % arXiv: \arxivtrue / ICDAR: \arxivfalse
\usepackage[T1]{fontenc}
\ifarxiv
  \usepackage{graphicx}
  \usepackage{CJKutf8}
\else
  \usepackage[dvipdfmx]{graphicx}
\fi
\usepackage{booktabs}
\usepackage{amsmath}
\usepackage{subcaption}
\begin{document}
\title{From Pen Strokes to Sleep States:\\Detecting Low-Recovery Days Using Sigma-Lognormal Handwriting Features}
\titlerunning{From Pen Strokes to Sleep States}
% If the paper title is too long for the running head, you can set
% an abbreviated paper title here
%
% === 著者・所属（arXiv版のみ表示） ===
\ifarxiv
  \author{Chisa Tanaka\inst{1} \and
  Andrew Vargo\inst{1} \and
  Anna Scius-Bertrand\inst{2} \and
  Andreas Fischer\inst{3} \and
  Koichi Kise\inst{1,4}}
  \authorrunning{C. Tanaka et al.}
  \institute{Osaka Metropolitan University, Osaka, Japan \and
  University of Applied Sciences and Arts Western Switzerland, Fribourg, Switzerland \and
  University of Fribourg, Fribourg, Switzerland \and
  German Research Center for Artificial Intelligence, Japan Laboratory, Osaka, Japan}
\fi
\maketitle              % typeset the header of the contribution
\begin{abstract}
% The abstract should briefly summarize the contents of the paper in
% 150--250 words.
While handwriting has traditionally been studied for character recognition and disease classification,
its potential to reflect day-to-day physiological fluctuations in healthy individuals remains unexplored.
This study examines whether daily variations in sleep-related recovery states
can be inferred from online handwriting dynamics.
We propose a personalized binary classification framework that detects low-recovery days
using features derived from the Sigma-Lognormal model,
which captures the neuromotor generation process of pen strokes.
In a 28-day in-the-wild study involving 13 university students,
handwriting was recorded three times daily,
and nocturnal cardiac indicators were measured using a wearable ring.
For each participant, the lowest (or highest) quartile of four sleep-related metrics---HRV,
lowest heart rate, average heart rate, and total sleep duration---defined the positive class.
Leave-One-Day-Out cross-validation showed that PR-AUC significantly exceeded the baseline (0.25)
for all four variables after FDR correction,
with the strongest performance observed for cardiac-related variables.
Importantly, classification performance did not differ significantly across task types or recording timings,
indicating that recovery-related signals are embedded in general movement dynamics.
These results demonstrate that subtle within-person autonomic recovery fluctuations
can be detected from everyday handwriting,
opening a new direction for non-invasive, device-independent health monitoring.
% TODO: 提出場所にそのまま載せることができるように1行でまとめたもの
% While handwriting has traditionally been studied for character recognition and disease classification, its potential to reflect day-to-day physiological fluctuations in healthy individuals remains unexplored. This study examines whether daily variations in sleep-related recovery states can be inferred from online handwriting dynamics. We propose a personalized binary classification framework that detects low-recovery days using features derived from the Sigma-Lognormal model, which captures the neuromotor generation process of pen strokes. In a 28-day in-the-wild study involving 13 university students, handwriting was recorded three times daily, and nocturnal cardiac indicators were measured using a wearable ring. For each participant, the lowest (or highest) quartile of four sleep-related metrics—HRV, lowest heart rate, average heart rate, and total sleep duration—defined the positive class. Leave-One-Day-Out cross-validation showed that PR-AUC significantly exceeded the baseline (0.25) for all four variables after FDR correction, with the strongest performance observed for cardiac-related variables. Importantly, classification performance did not differ significantly across task types or recording timings, indicating that recovery-related signals are embedded in general movement dynamics. These results demonstrate that subtle within-person autonomic recovery fluctuations can be detected from everyday handwriting, opening a new direction for non-invasive, device-independent health monitoring.
% TODO(田中): キーワードを先生に確認しに行く
\end{abstract}
%
% TODO(メモだから本文の内容を確認した上で対応): Low-Recovery Daysの意味をイントロに入れる

\section{Introduction}

% 手書き側から書く「オンラインの手書きデータ, 文字の情報だけでなくいろんな情報が含まれている,
% ADHD, 性別, 本人, アカデミックパフォーマンス, とかいろいろされている．他にはないの？
% この論文では別の要素を．個人の健康とか体調well being wellness, 睡眠とHRに着目する．
% HRからストレスに．先行研究は，HRは無いし，睡眠もない，"新たな領域に挑戦する"」
% 「手書きから文字認識がメインだった．今はいろいろできる．それが全てか？
% 他にも含まれている可能性がある．睡眠とHR．リアルタイムではないが，ストレスが」

% 1. オンライン手書きデータの豊かさ
Online handwriting data acquired via digital pens and tablet devices contain rich information
related to the writer's motor control, including not only character shape information
but also writing speed, pen pressure, and inter-stroke time intervals.
While the primary application of handwriting analysis has been character recognition,
recent research has leveraged such kinematic features
to estimate various attributes and states of the writer.

% 2. これまでの応用（ADHD, 神経変性疾患, 加齢, 性別, 認証...）
For example, ADHD classification using handwriting features~\cite{ADHD},
early diagnosis of neurodegenerative diseases from handwritten signatures~\cite{pirlo2015neurodegeneration},
analysis of changes in motor control ability associated with aging~\cite{plamondon2013lognormal},
% オンライン手書きからの性別推定~\cite{sesanogueras2016gender}，
and prediction of cognitive performance using dynamic features from digital pens~\cite{prange2022cognitive}
have been reported across diverse applications.
These studies demonstrate that handwriting is not merely a means of conveying written information,
but a rich source of information reflecting the writer's neurological and cognitive characteristics.

% 3. 新たな領域への挑戦 — 健康・睡眠・HR
However, it remains unclear whether handwriting contains additional information beyond these applications.
The aforementioned studies primarily target the identification of pathological conditions or developmental stages;
estimating day-to-day variations in physical condition among healthy individuals remains an unexplored area.
This study explores the new domain of estimating personal health status from handwriting.
Specifically, we focus on sleep duration and sleep-related cardiac indicators.
Heart rate variability (HRV) and heart rate during sleep are objective indicators
reflecting the recovery state of the autonomic nervous system~\cite{tobaldini2013hrv},
and decreases in HRV and increases in resting heart rate have been reported to be associated with accumulated stress~\cite{kim2018stress}.
Handwriting is a form of fine motor control,
and it has been reported that poor sleep quality impairs fine motor control~\cite{venevtseva2024sleep}.
Therefore, changes in sleep quality may be reflected in daytime handwriting patterns.

% 4. ウェアラブルデバイスの限界 → 手書きによる代替
Currently, cardiac data during sleep can be obtained using wearable devices such as smartwatches and smart rings;
however, these devices pose several challenges for accessibility.
Besides being expensive, they require that users continuously wear the device, interact with it, and frequently maintain it.
This daily interaction can lead to high levels of device abandonment~\cite{lazar2015abandon}.
If sleep quality could be estimated from everyday handwriting activities,
non-invasive sleep monitoring independent of wearable devices would become possible.
This could make sleep quality estimation accessible for a wider audience,
such as children who already use tablets at school.

% 5. 本研究の提案
This study proposes a binary classification framework that detects days with poor sleep quality (low-recovery days)
from handwriting features based on the Sigma-Lognormal model.
The main contributions of this study are as follows:
\begin{itemize}
  \item We introduce a previously unexplored application domain: estimating sleep quality from handwriting features.
  \item Using 28-day in-the-wild data from 13 participants evaluated with person-dependent leave-one-day-out cross-validation,
        we confirmed that classification performance significantly exceeded the baseline
        for all four sleep variables (after false discovery rate (FDR) correction).
  \item No significant differences in classification performance were observed across five handwriting task types
        or three daily session timings (after FDR correction),
        suggesting that the proposed approach is largely robust to task and timing variations within the evaluated conditions.
\end{itemize}
These results suggest the feasibility of non-invasive sleep monitoring
through everyday handwriting activities in educational settings where tablet devices are widely adopted.

\section{Related Work}

% TODO(田中): 引用の書き方が正しいか確認する

\subsection{State Estimation via Handwriting Analysis}

Research has been conducted on estimating the writer's state and characteristics
from kinematic features of online handwriting data.
Faci et al.~\cite{ADHD} classified 12 children with ADHD and 12 typically developing children
using handwriting features extracted from the Sigma-Lognormal model,
achieving a classification accuracy of 91.67\%.
Pirlo et al.~\cite{pirlo2015neurodegeneration} used dynamic parameters
similarly extracted from the Sigma-Lognormal model
to attempt early diagnosis of neurodegenerative diseases such as Alzheimer's disease from handwritten signatures,
reporting an error rate of 3\%.
Furthermore, Plamondon et al.~\cite{plamondon2013lognormal} analyzed
the process by which motor control ability changes with learning, skill acquisition, and aging
within the theoretical framework of the Sigma-Lognormal model,
demonstrating that variations in model parameters reflect the state of the neuromotor system.

These studies demonstrate that handwriting features based on the Sigma-Lognormal model
can capture neuromotor changes attributable to diseases and developmental stages.
However, the targets are all pathological conditions or long-term developmental changes,
and no studies have been found that estimate day-to-day variations in physical condition
(such as fatigue and sleep quality) among healthy individuals from handwriting.

\subsection{Sleep and Handwriting}

It has been reported that sleep quality affects daytime motor control.
Venevtseva et al.~\cite{venevtseva2024sleep} demonstrated in a study of university students
that poor sleep quality impairs fine motor skills.

As a study directly investigating the relationship between handwriting and sleep,
Jasper et al.~\cite{jasper2009circadian} analyzed handwriting kinematics
(writing speed, fluency, character size, etc.) every three hours
in nine healthy young men under a 40-hour sleep deprivation protocol (laboratory setting).
While a circadian rhythm was confirmed in writing speed,
writing fluency did not change significantly throughout all sessions,
and the effects of sleep deprivation were not reflected in the daytime kinematics on the second day.
This was attributed to the robust maintenance of basic handwriting automatization in skilled writers.
However, the features used in this study were basic kinematics such as writing speed and fluency,
and Sigma-Lognormal model parameters that more deeply decompose the motor generation process were not employed.
Moreover, the study by Jasper et al.\ examined the effects of acute sleep deprivation
in a controlled laboratory setting,
and the relationship with natural variations in sleep quality under in-the-wild conditions has not been verified.

The above findings suggest the possibility that changes in sleep quality affect handwriting;
however, no study has predicted sleep-related indicators from handwriting features.
This study attempts to classify sleep indicators under in-the-wild conditions
using finer-grained features based on the Sigma-Lognormal model.

% === tanaka2025jsai（自己実験）の記載：ダブルブラインド確認後に検討 ===
% 先行研究~\cite{tanaka2025jsai}において，筆者自身を被験者とする自己実験により，
% 手書き特徴量と睡眠データの関係性を調査した．
% この研究では，起床後・昼食後・就寝前の1日3回，
% 直線，意味のない文字列，ポジティブ・ネガティブなフレーズ，本の引用の5種類の手書きタスクを実施し，
% シグマログノーマルモデルで特徴量を抽出した．
% Oura Ringで取得した睡眠時間を目的変数として，
% AICに基づく線形モデルで解析を行った結果，
% 全ての睡眠変数について高い適合度が得られた．
% しかし，この結果は1名の自己実験に基づくものであり，予測ではなく推定である．

\begin{table}[t]
  \centering
  \caption{Five basic features extracted from the Sigma-Lognormal model. The mean and standard deviation of each feature are computed, yielding ten features in total.}
  \label{tab:features}
  \small
  \begin{tabular}{@{}cll@{}}
    \toprule
    Feature & Description & Type \\
    \midrule
    $D$ & Amplitude of motor command & Model parameter \\
    $t_0$ & Onset time of motor command & Model parameter \\
    nblog & Number of lognormal distributions & Reconstruction index \\
    SNR & Signal-to-noise ratio & Reconstruction index \\
    SNR/nblog & Reconstruction accuracy per distribution & Reconstruction index \\
    \bottomrule
  \end{tabular}
\end{table}

\begin{figure}[t]
    \centering
    \fbox{
    \includegraphics[width=0.4\linewidth]{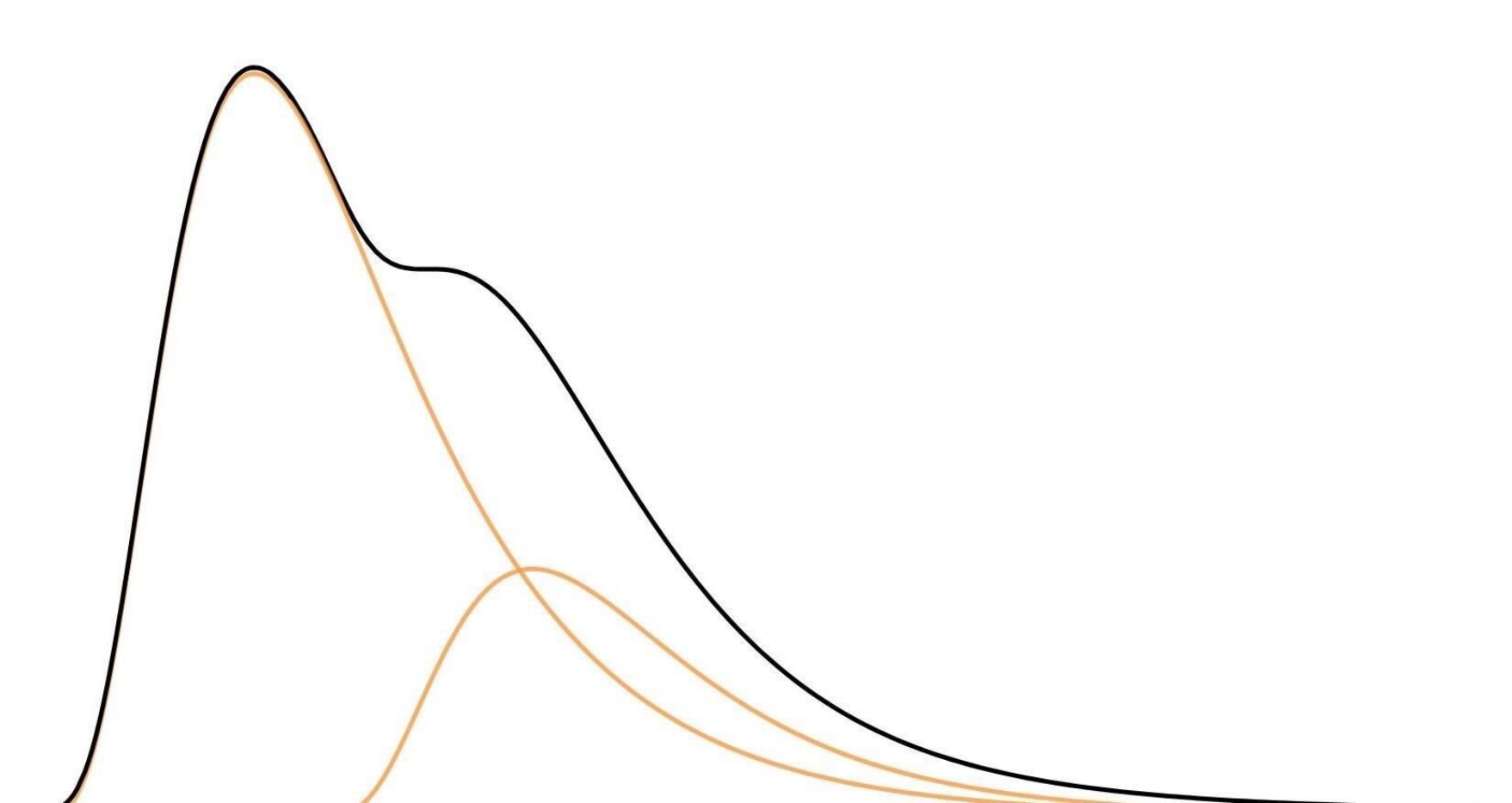}
    }
    \caption{Illustration of the Sigma-Lognormal model.\\
    The velocity profile (black line) is described as a sum of lognormal distributions (yellow lines).}
    \label{fig:sigma}
\end{figure}

\section{Proposed Method}

\subsection{Sigma-Lognormal Model}
\label{subsec:sigma}

Handwriting is a form of human movement and can be described mathematically by motor theory~\cite{plamondon2021lognormality}.
The Sigma-Lognormal model~\cite{oreilly2009sigma} describes the velocity profile of a single stroke as a superposition of lognormal distributions (Fig.~\ref{fig:sigma}), and is expressed by the following equation:
\begin{equation}
  \label{eq:sigma}
  \vec{v}(t)
  = \sum_{i=1}^N D_i
  \begin{bmatrix}
    \cos(\theta_i(t))\\
    \sin(\theta_i(t))
  \end{bmatrix}
  \Lambda_i(t;t_{0i}, \mu_i, \sigma_i^2),
  \quad N \ge 2
\end{equation}
where $D_i$ is the amplitude of the motor command sent from the brain at time $t_{0i}$,
and $\Lambda_i$ is the motor response represented by a lognormal distribution.

In our method, parameter extraction is performed using the implementation provided by Lai et al.~\cite{lai2022synsig2vec}\footnote{\url{https://github.com/LaiSongxuan/SynSig2Vec}},
which is based on the method of O'Reilly and Plamondon~\cite{oreilly2009sigma}.
On our dataset, we achieved a mean signal-to-noise ratio (SNR) of $24.1 \pm 4.4$~dB, indicating good reconstruction quality of the Sigma-Lognormal model.
From this model, we extract five basic features shown in Table~\ref{tab:features}.
$D$ and $t_0$ are parameters of the lognormal distributions constituting the Sigma-Lognormal model,
while nblog, SNR, and SNR/nblog are indicators of the goodness of fit between the original velocity profile and the reconstructed one.
Since each handwriting task contains multiple strokes,
we compute the mean and standard deviation of each feature across the entire task, yielding ten features in total.

\begin{figure}[t]
  \centering
  \begin{subfigure}[c]{0.48\textwidth}
    \centering
    \begin{subfigure}[b]{\linewidth}
      \centering
      \includegraphics[width=0.5\linewidth]{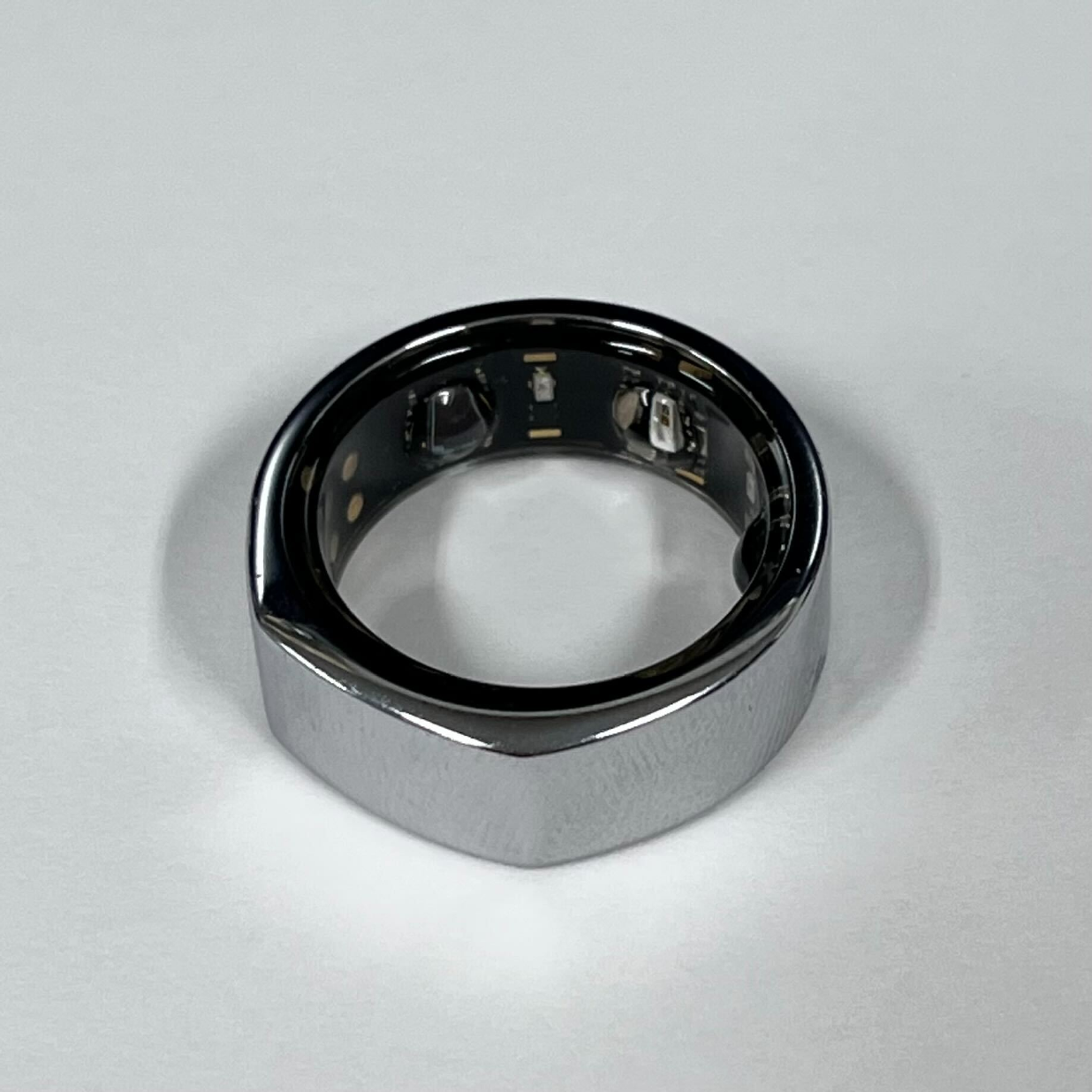}
      \caption*{Generation 3}
    \end{subfigure}
    \\[2mm]
    \begin{subfigure}[b]{\linewidth}
      \centering
      \includegraphics[width=0.5\linewidth]{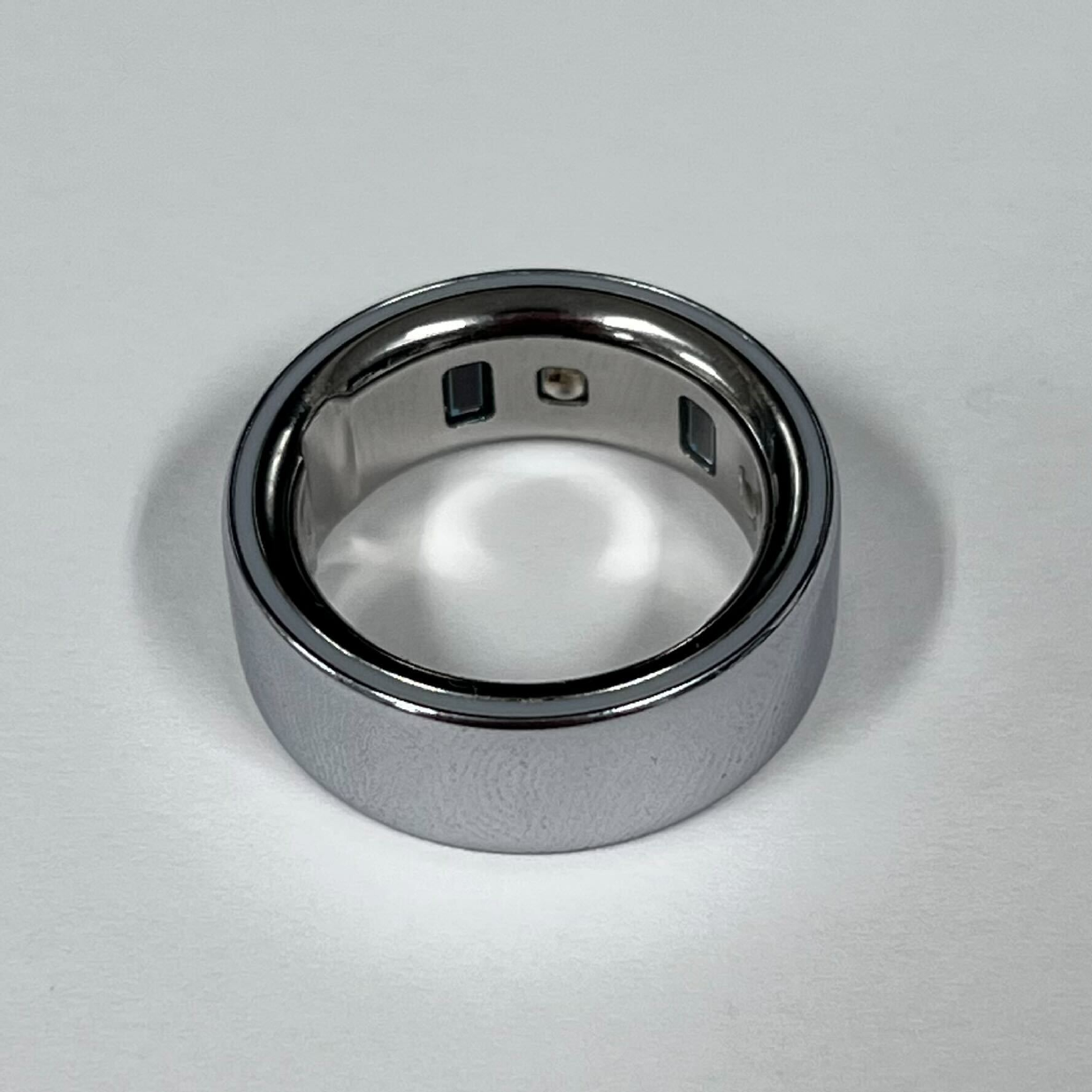}
      \caption*{Generation 4}
    \end{subfigure}
    \caption{Oura Ring}
    \label{fig:oura_proposed}
  \end{subfigure}
  \hfill
  \begin{subfigure}[c]{0.45\textwidth}
    \centering
    \includegraphics[width=\linewidth]{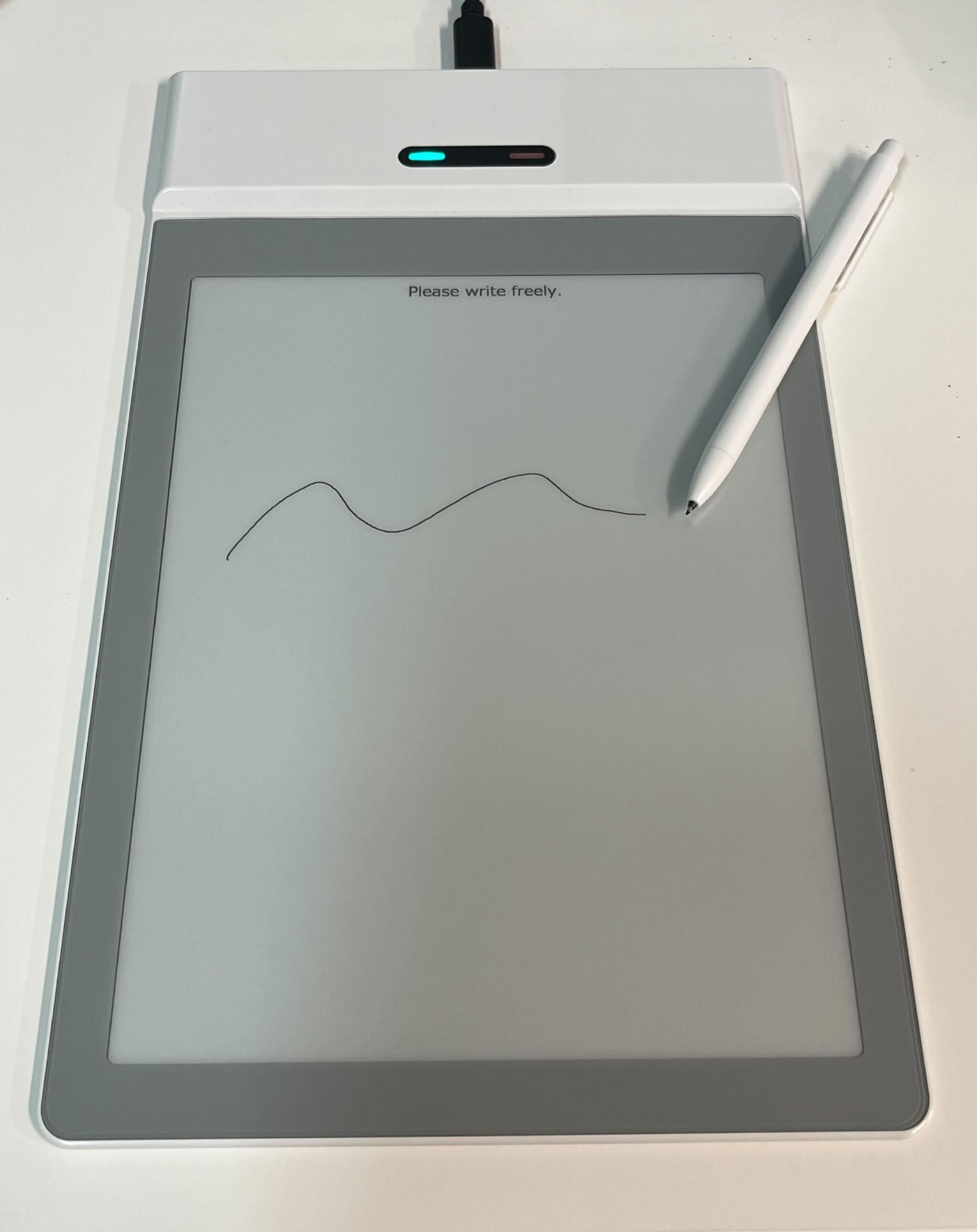}
%    \caption{IoT Paper}
    \caption{Wacom IoT Paper}
    \label{fig:iot_proposed}
  \end{subfigure}
  \caption{Devices used for data collection}
  \label{fig:devices}
\end{figure}

\subsection{Sleep Indicators}
\label{subsec:sleep_indicators}

We define four target variables representing sleep quality,
all derived from data recorded by the Oura Ring (Fig.~\ref{fig:devices}~(a)) during sleep.
Total sleep duration is the total time spent asleep
and serves as the most fundamental measure of sleep quantity.
Average heart rate variability (Avg HRV) reflects the recovery state of the autonomic nervous system,
particularly parasympathetic activity;
higher HRV during sleep indicates better physiological recovery.
Lowest heart rate (Lowest HR) is the minimum heart rate observed during sleep,
reflecting the deepest resting state.
Average heart rate (Avg HR) is the mean heart rate during sleep;
elevated resting heart rate has been associated with accumulated stress~\cite{kim2018stress}.

These four variables were selected because they are based on raw sensor measurements (optical heart rate sensor)
rather than device-specific algorithmic estimates such as sleep stage classification (deep, light, and REM sleep),
which may vary across device manufacturers and firmware versions.

\subsection{Binary Classification Framework}
\label{subsec:classification}

In this study, rather than predicting sleep-related variables,
we address a binary classification problem of detecting ``days with poor sleep quality.''
For each target variable, the first quartile (Q25) of each user is used as the threshold,
and samples below (or above) the threshold are defined as the ``problematic'' class (positive).
The threshold direction is set based on the nature of each variable:
for sleep duration and HRV, lower values indicate problems, so the lower direction is used (below Q25 is positive);
for heart rate, higher values indicate problems, so the upper direction is used (above Q75 is positive).
This quartile-based thresholding strategy was adopted to standardize the positive class proportion across individuals
while focusing on relatively extreme recovery states.
A separate binary classifier is trained and evaluated independently for each of the four target variables;
thus, the four classification tasks do not interact with each other.
% TODO: 文字数に余裕があれば引用を追加する
This results in approximately 25\% of samples being positive, creating an imbalanced classification problem.

% 速度可視化 + インストラクション
\begin{figure}[p]
  \centering
  \includegraphics[width=0.3\columnwidth]{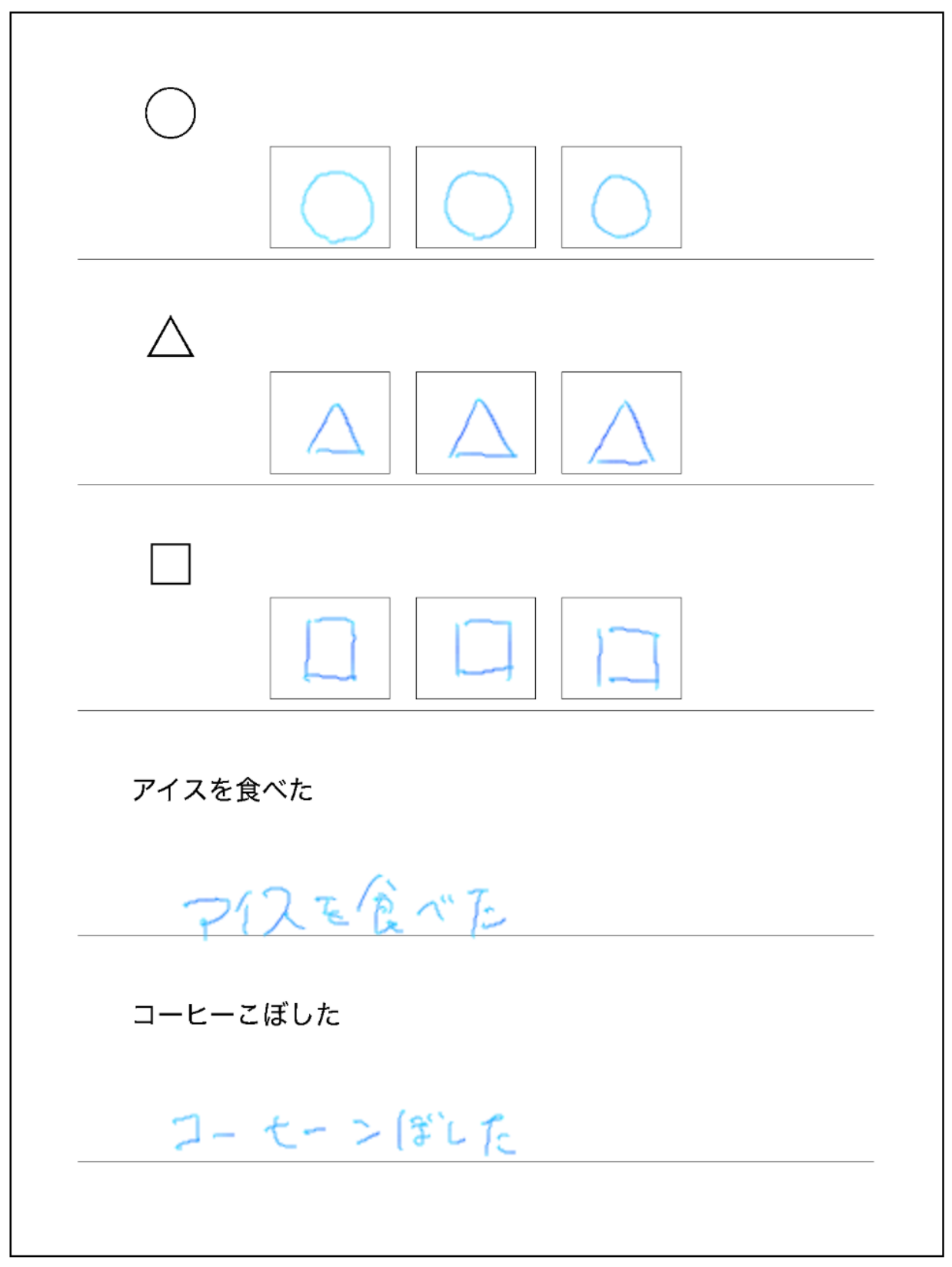}
  \hfill
  \includegraphics[width=0.3\columnwidth]{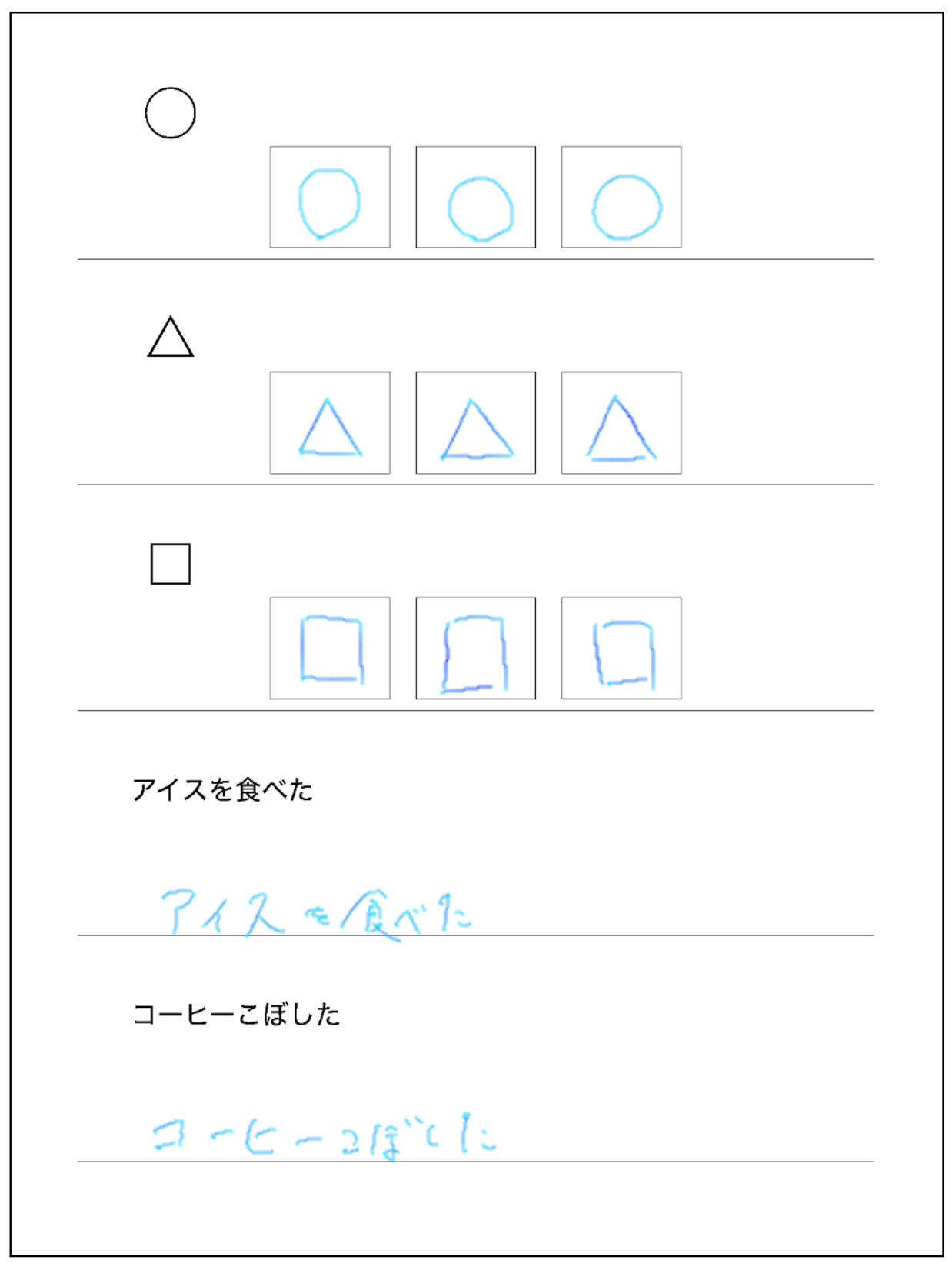}
  \hfill
  \includegraphics[width=0.3\columnwidth]{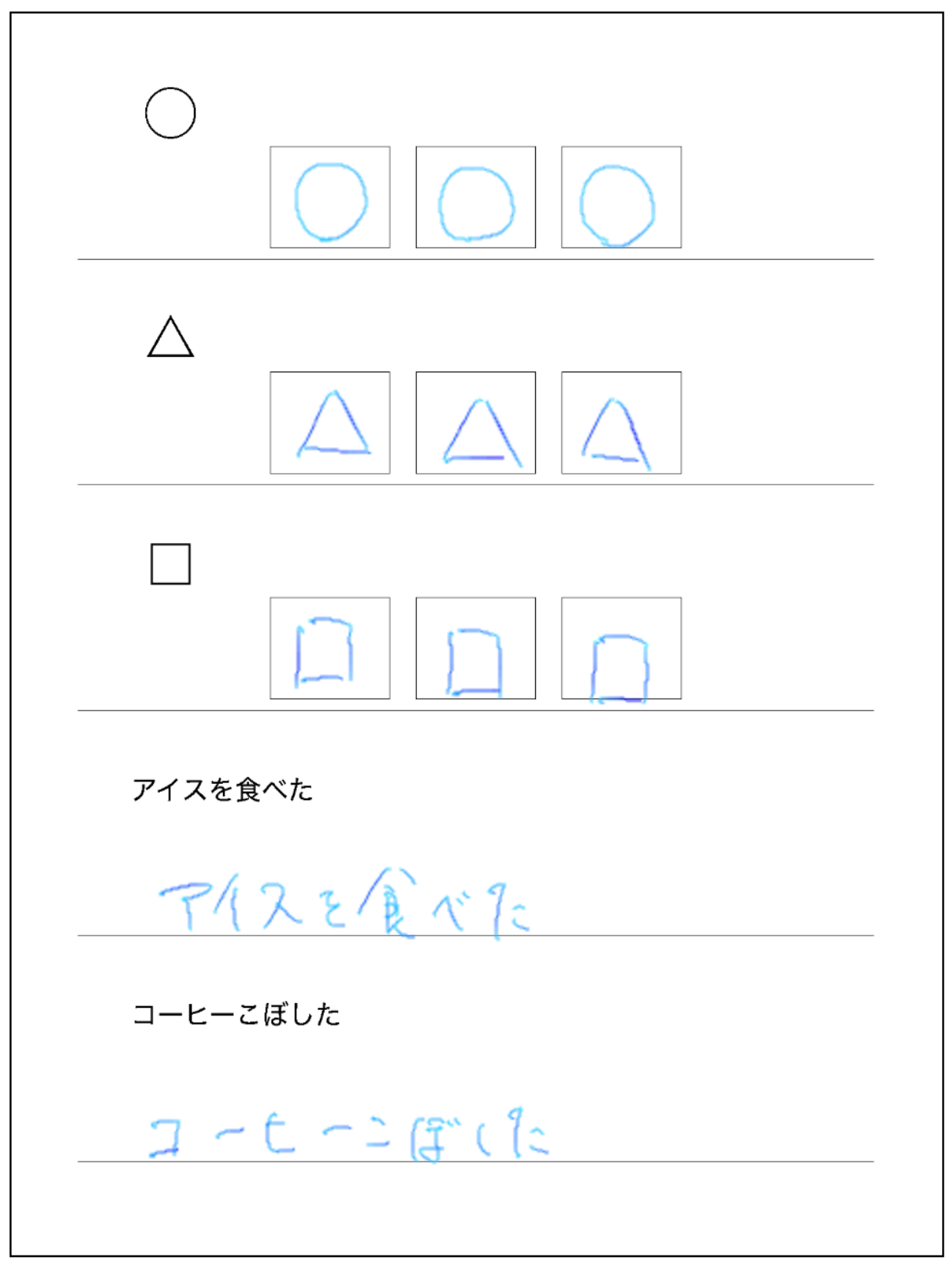}\\[2pt]
  \includegraphics[width=0.3\columnwidth]{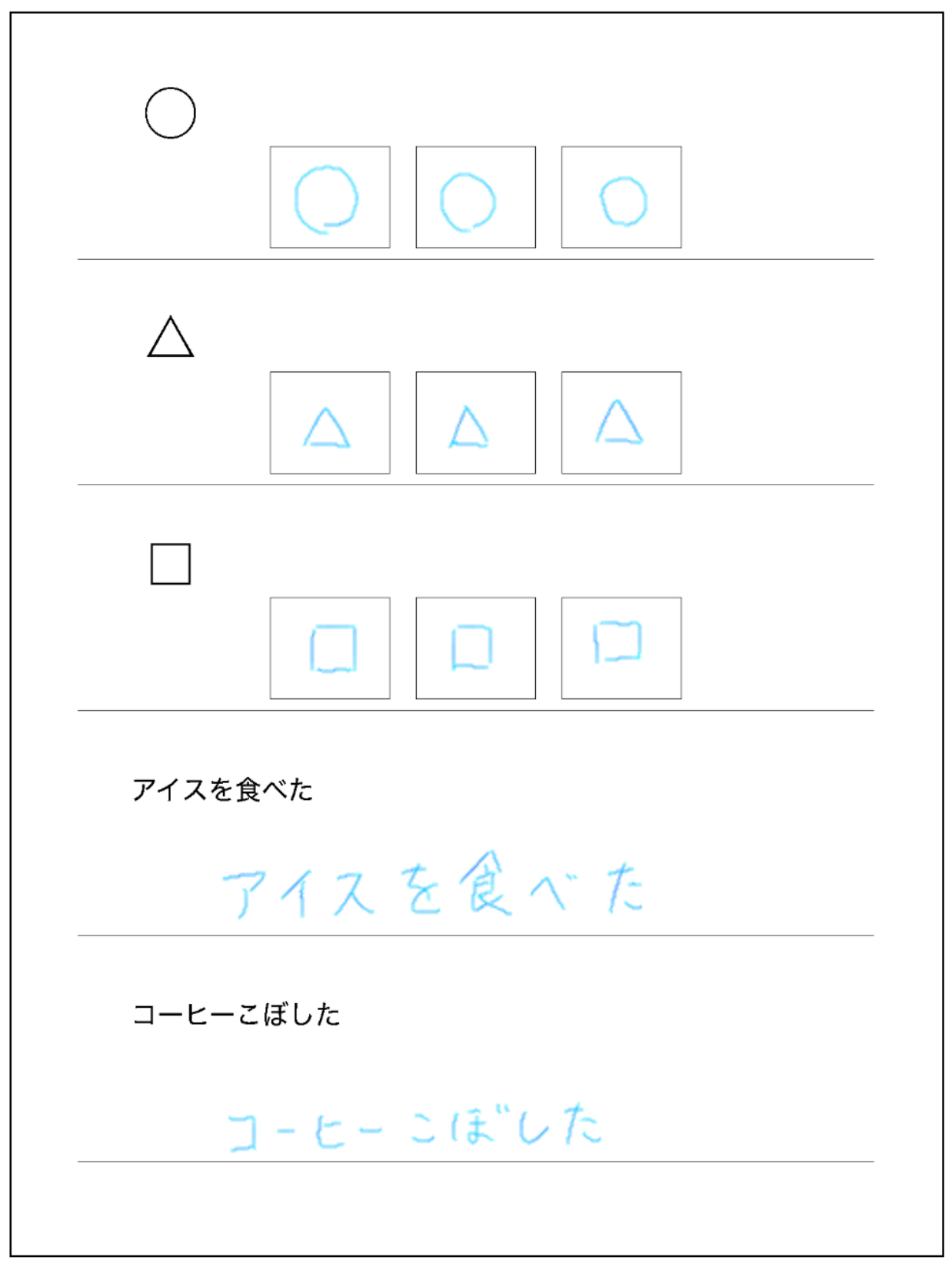}
  \hfill
  \includegraphics[width=0.3\columnwidth]{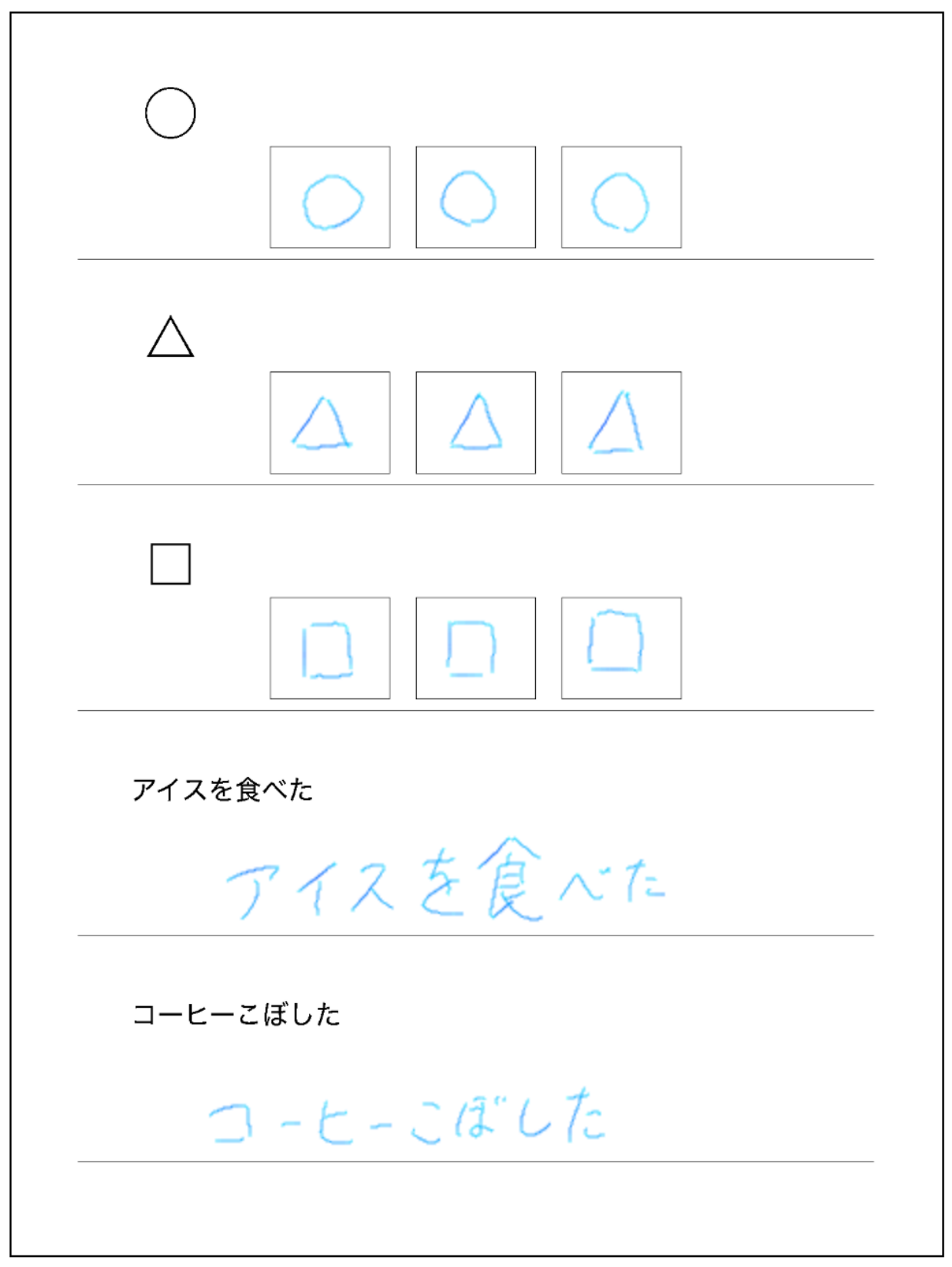}
  \hfill
  \includegraphics[width=0.3\columnwidth]{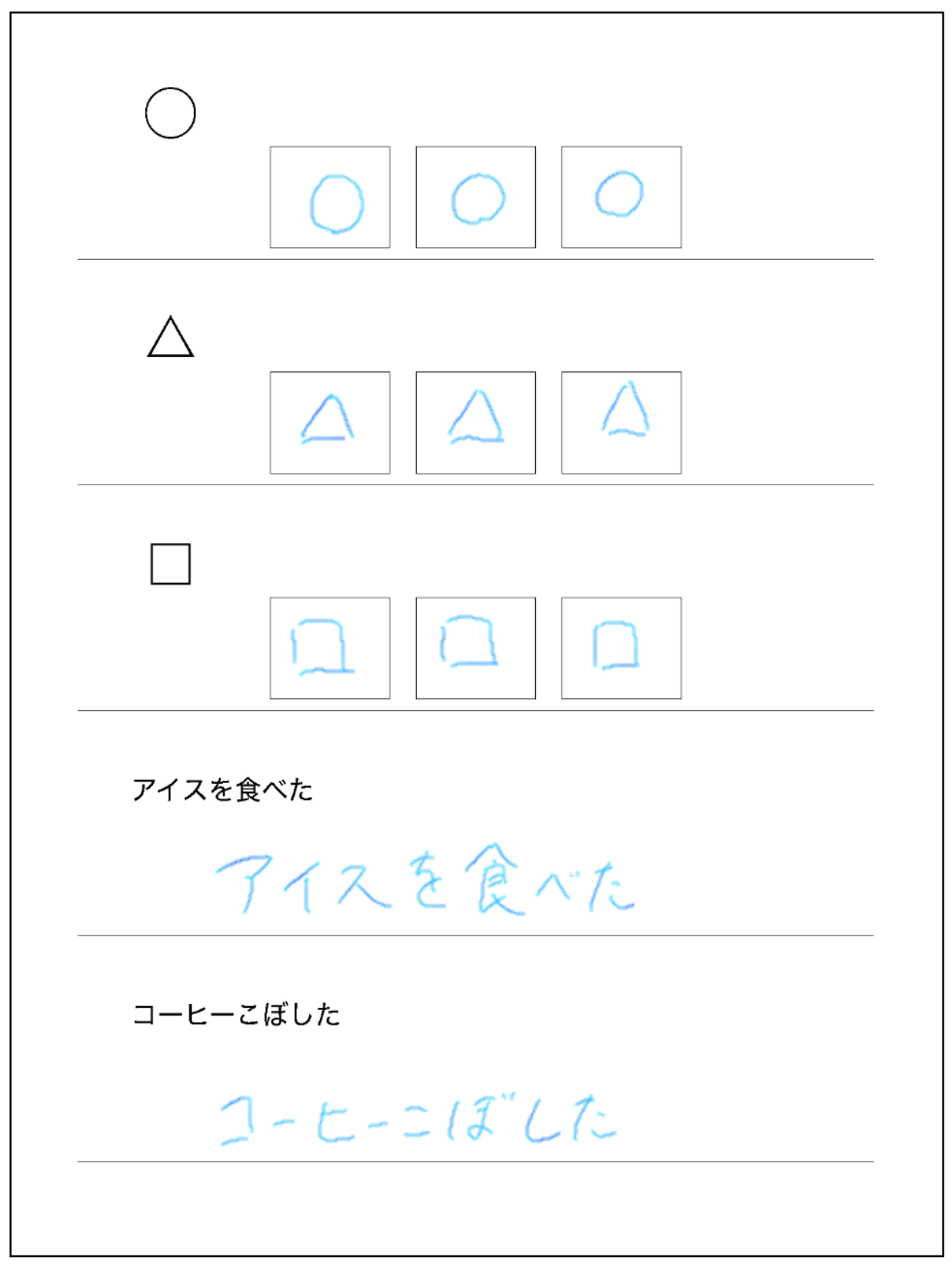}\\[2pt]
  \includegraphics[width=0.3\columnwidth]{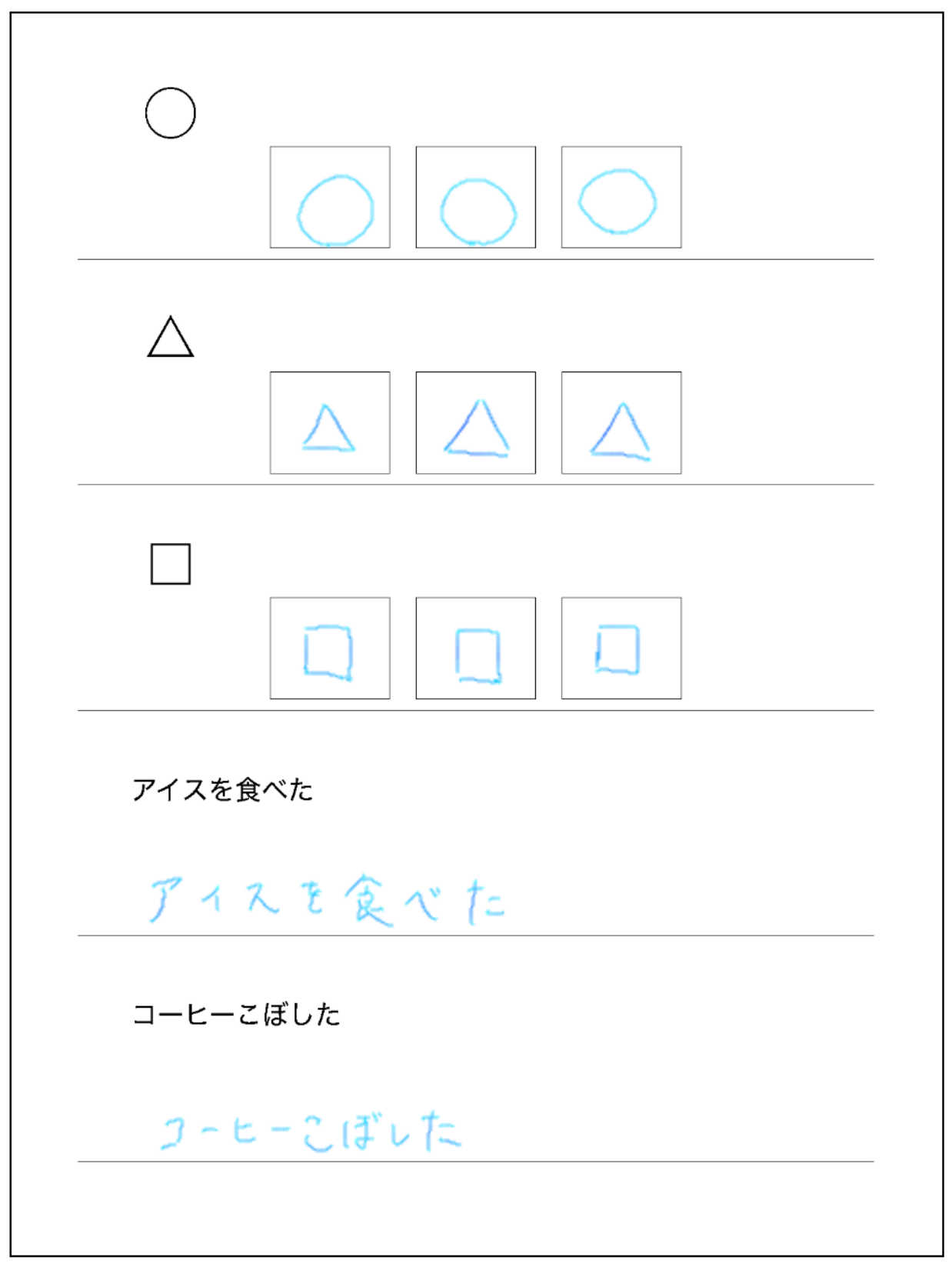}
  \hfill
  \includegraphics[width=0.3\columnwidth]{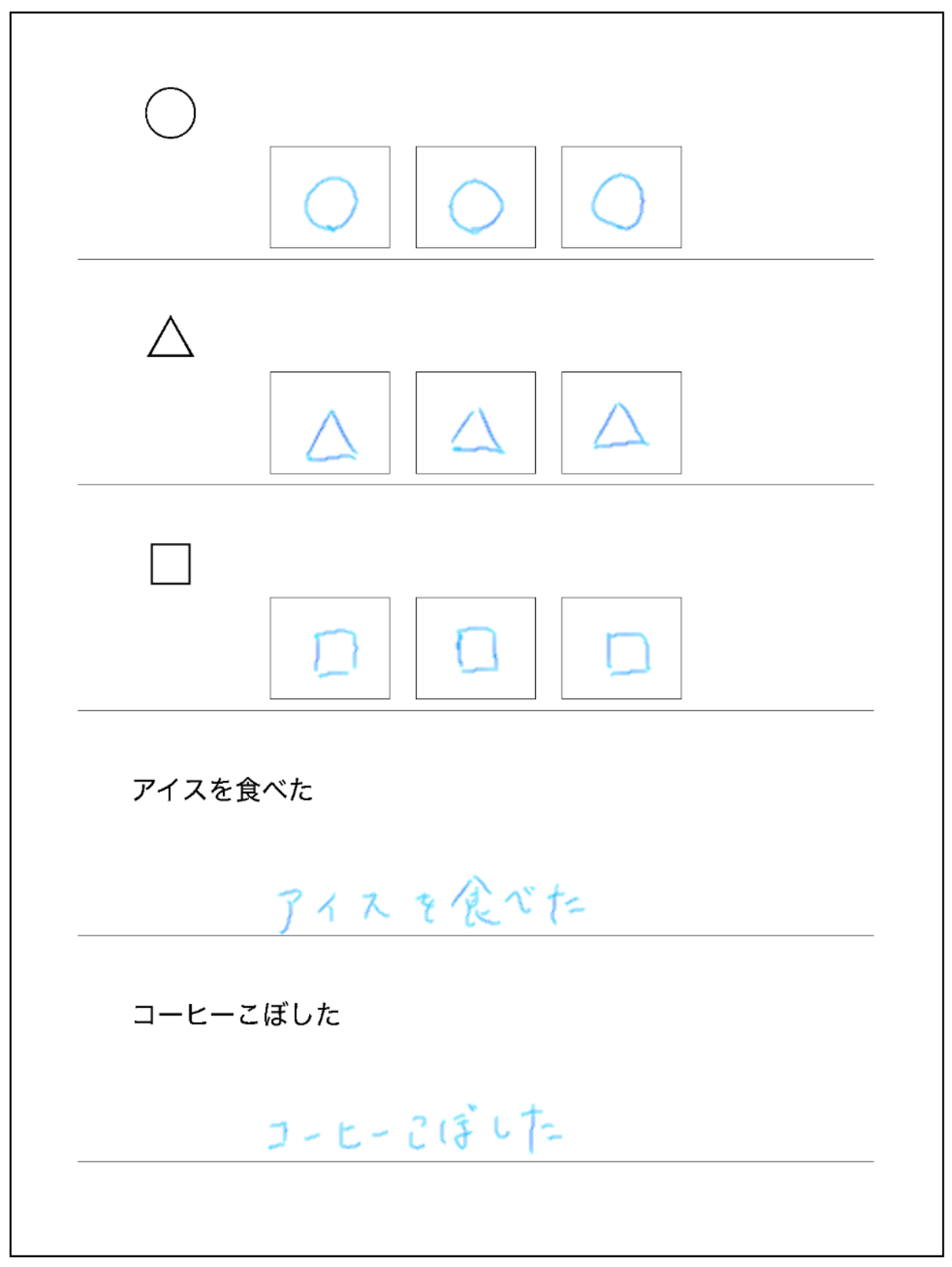}
  \hfill
  \includegraphics[width=0.3\columnwidth]{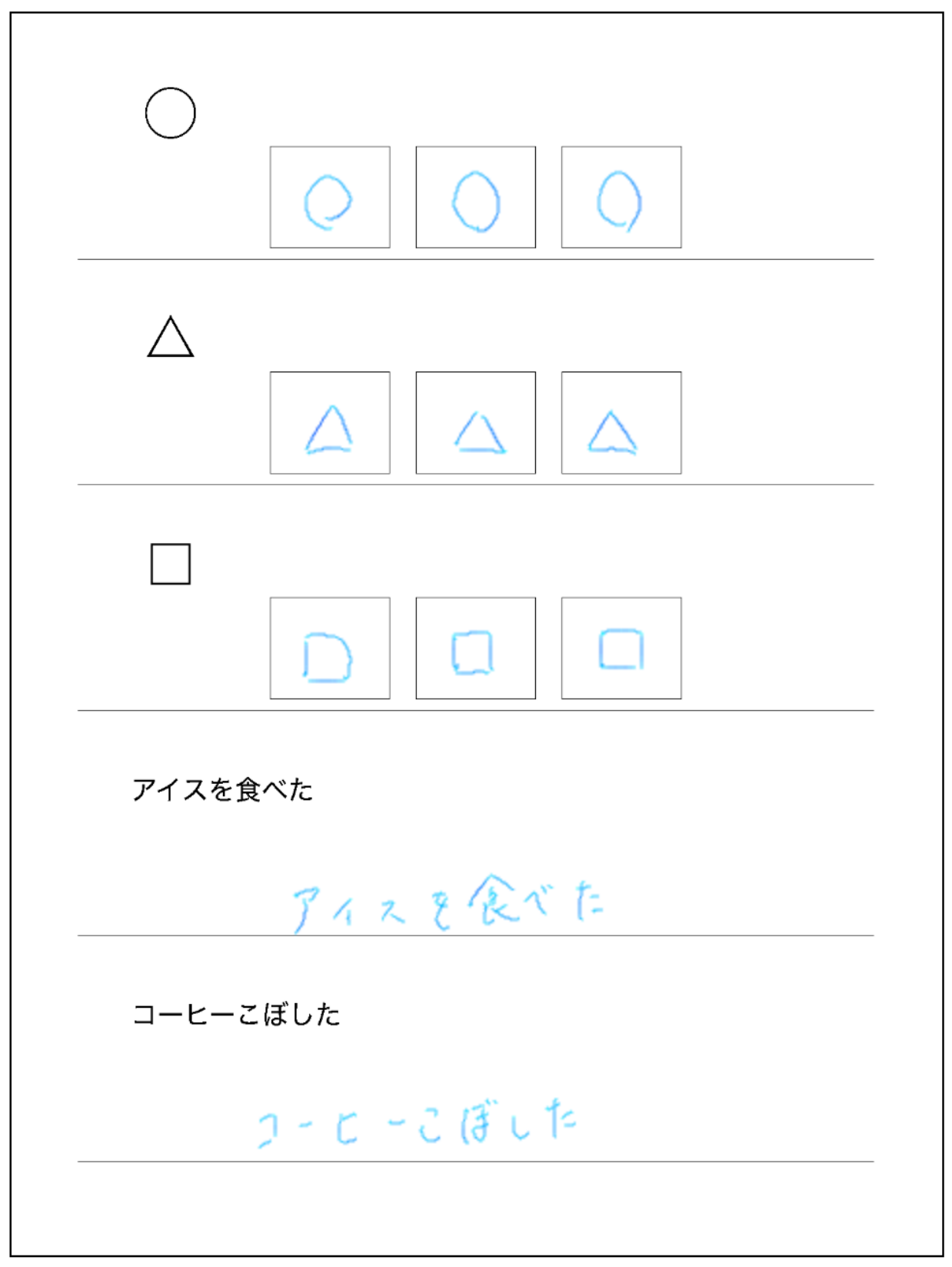}\\[4pt]
  \includegraphics[width=0.6\columnwidth]{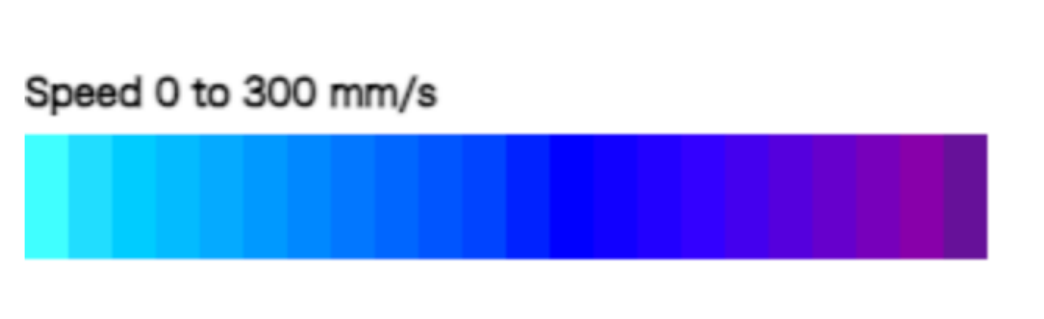}
  \caption{Visualization of writing speed for three participants (P1, P7, P13). Each row represents one participant, and each column shows data from days 1, 14, and 28 of the experiment. Color indicates stroke speed (0--300\,mm/s).}
  \label{fig:handwriting_speed}
\end{figure}

\begin{figure}[!ht]
  \centering
  \begin{subfigure}[b]{0.25\columnwidth}
    \centering
    \includegraphics[width=0.7\linewidth]{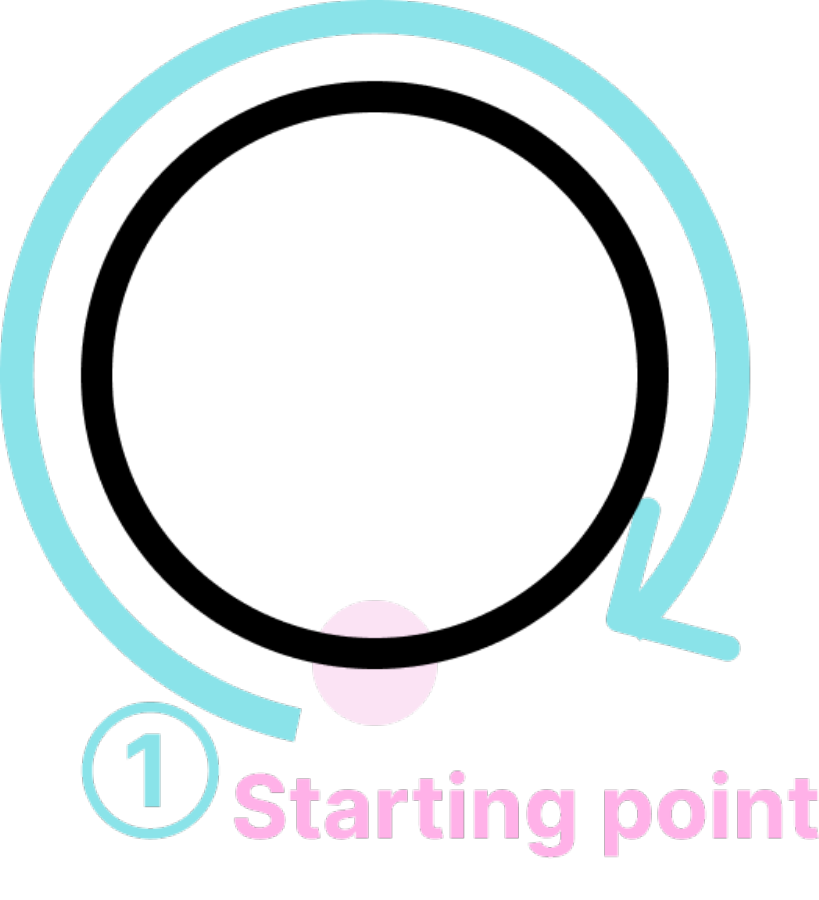}
    \caption{Circle}
    \label{fig:handwriting_inst_circle}
  \end{subfigure}
  \hfill
  \begin{subfigure}[b]{0.25\columnwidth}
    \centering
    \includegraphics[width=0.7\linewidth]{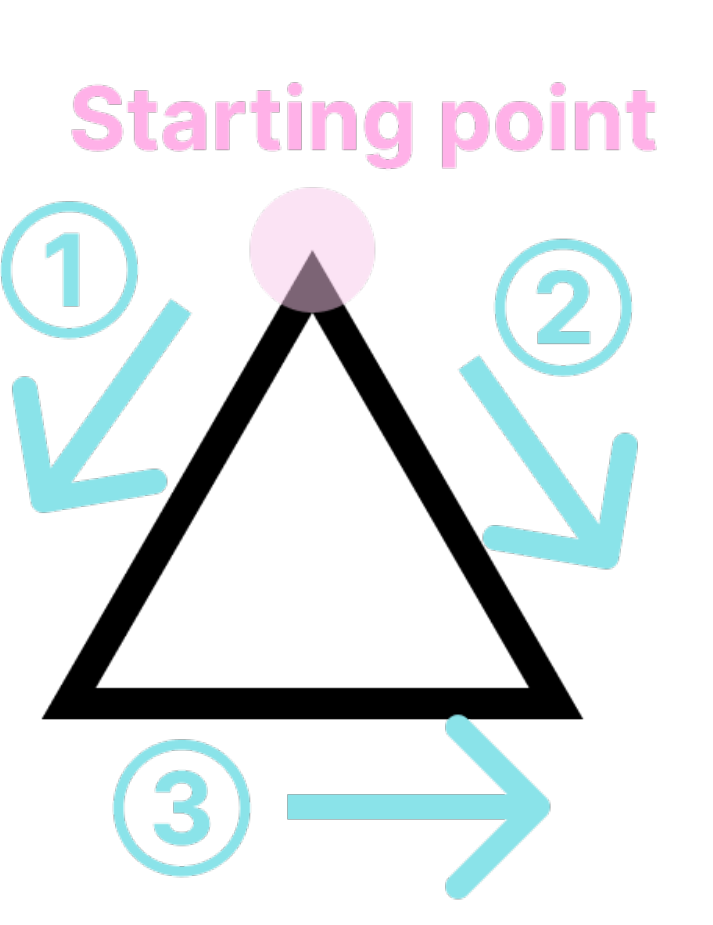}
    \caption{Triangle}
    \label{fig:handwriting_inst_triangle}
  \end{subfigure}
  \hfill
  \begin{subfigure}[b]{0.25\columnwidth}
    \centering
    \includegraphics[width=0.95\linewidth]{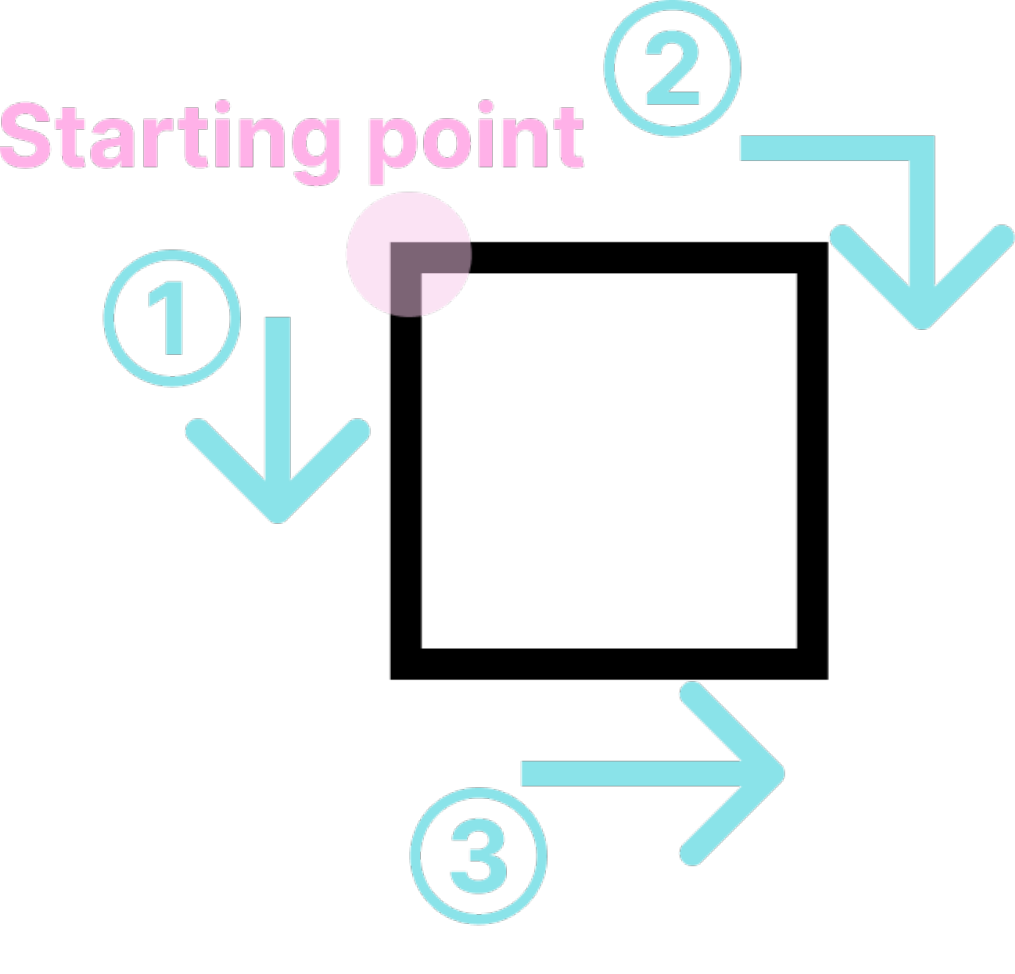}
    \caption{Square}
    \label{fig:handwriting_inst_square}
  \end{subfigure}
  \caption{Stroke order instructions for handwriting tasks}
  \label{fig:handwriting_inst}
\end{figure}

A Random Forest (RF) is used as the classifier.
RF is a type of ensemble learning that can capture nonlinear interactions among features
and exhibits relatively stable performance even with a small number of samples.
Preliminary experiments with linear classifiers showed comparable or lower performance;
therefore, RF was selected for its robustness in small-sample nonlinear settings.

\section{Dataset}
\label{sec:dataset}

% === 分量調整メモ（2026-02-25 議論） ===
% 現状: 最小限の記載（データ除外 + 睡眠統計量を本文中に記載）
% スペースに余裕がある場合に追加検討する項目:
%   - 睡眠変数の基本統計量の表（Table化）
%   - 図形タスクのストローク数が実験後半で変化した参加者がいた観察
%     → 特徴量はタスク全体の平均・SDなので分析への影響は限定的だが，
%       学習効果・慣れの議論としてDiscussionに1文入れる余地あり
%   - 睡眠データの分布図（ヒストグラム等）
%     → スペース消費大，本題との距離があるため優先度低
%   - デバイス写真のサイズ・レイアウト調整で節約可能
% ============================================

The dataset used in this study was newly collected for the purpose of this research and consists of handwriting data and sleep data.

Handwriting data were collected using a tablet device called IoT Paper (Fig.~\ref{fig:devices}~(b)) with a sampling frequency of 480~Hz, provided for research purposes by Wacom Co., Ltd.
A dedicated electronic pen allows natural handwriting similar to writing on paper.
Online handwriting data including $x$- and $y$-coordinates, pen pressure, tilt, and pen-to-screen distance were recorded in InkML format.
% TODO(田中): IoT Paperから取得できるデータについて(市販されておらず，研究用のデバイスで情報が公開されていないため，田中がデータを確認して記載する必要がある)

Sleep data were measured using Oura Ring (Fig.~\ref{fig:devices}~(a)), a ring-type wearable device.
Oura Ring records heart rate, heart rate variability (HRV), and sleep stages (deep sleep, light sleep, and REM sleep) during sleep using optical sensors.
The validity of sleep measurement has been confirmed through comparison with polysomnography~\cite{oura_validation},
and the highest accuracy among commercially available wearable devices has been reported, particularly for resting heart rate and HRV measurements~\cite{dial2025oura}.
In this experiment, two generations (Gen~3 and Gen~4) of Oura Ring were used.
A previous study~\cite{dial2025oura} has reported that both Gen~3 and Gen~4 achieve accuracy comparable to medical-grade electrocardiography for measuring sleep-related cardiac indicators,
with no significant differences between generations.
Measurement data were obtained in CSV format through the Oura API v2\footnote{\url{https://cloud.ouraring.com/v2/docs}}.

% 実際に取得した手書きデータの指示
The data collection period was set to 28 days, and we adopted a format in which the experiment continued until 28 days of data were collected.
Seventeen university students participated in the experiment.
Of the 17 participants, 4 were female and 13 were male.
All 17 participants were first-time users of the IoT Paper.
Handwriting data were collected three times per day: after waking, after lunch, and before bedtime.
The tasks consisted of shape drawing tasks in which participants drew circles, triangles, and squares three times each per session,
and phrase writing tasks in which participants wrote a positive phrase
(\ifarxiv``\begin{CJK}{UTF8}{min}アイスを食べた\end{CJK}''\else``アイスを食べた''\fi(I ate ice cream))
and a negative phrase
(\ifarxiv``\begin{CJK}{UTF8}{min}コーヒーこぼした\end{CJK}''\else``コーヒーこぼした''\fi(I spilled coffee)) each.
Figure~\ref{fig:handwriting_speed} shows the visualization of writing speed for three participants on days 1, 14, and 28 of the experiment.
Stroke speed patterns vary across days even within the same participant.
For participants who ate only two meals per day,
they were instructed to perform the after-lunch task after their first meal (provided that at least two hours had elapsed since waking).
The same tasks were performed at all timings throughout the 28-day period.
For shape drawing tasks, stroke order instructions were provided before the experiment as shown in Fig.~\ref{fig:handwriting_inst}~(a) through Fig.~\ref{fig:handwriting_inst}~(c), encouraging participants to draw shapes in the instructed order.
% \ref{fig:task_results}~(a)
% 実験参加者全員に対し，書き順の説明の際に，手書きタスクに取り組む際に毎回書き順の例示が必要であるかを確認し，
% 17名全員が不要と回答したため，実験期間中の書き順の例示は行わなかった．
However, some participants were observed to have changed the stroke order for shape tasks in the latter half of the experiment, resulting in changes in the number of strokes.

Four participants had been continuously wearing the Oura Ring since before the experiment began.
For new users (13 participants), a break-in wearing period of at least one week was established before the experiment for device calibration.
To capture more natural variations in sleep quality, daytime wearing was optional,
while continuous wearing from bedtime to waking was mandatory.

For each timing (after waking, after lunch, and before bedtime) independently,
the experiment was continued until 28 days of paired handwriting and sleep data were collected.
% にわたり手書きデータと睡眠データの取得を行った．
Of the 17 participants, 4 were excluded from the analysis
because they frequently forgot to perform tasks and could not collect 28 days of complete data.
% データ取得が不完全であったため
Data from 13 participants (28 days $\times$ 3 timings $\times$ 5 tasks = 420 samples per participant) were used.
Although this results in 420 handwriting samples per participant, the sleep label is shared within each day.

% 実験参加者ごとのターゲット変数の分布（箱ひげ図 + ストリッププロット）
% 参加者はTotal Sleepの中央値が小さい順にP1〜P13とラベル付け
\begin{figure}[p]
  \centering
  \includegraphics[width=\columnwidth]{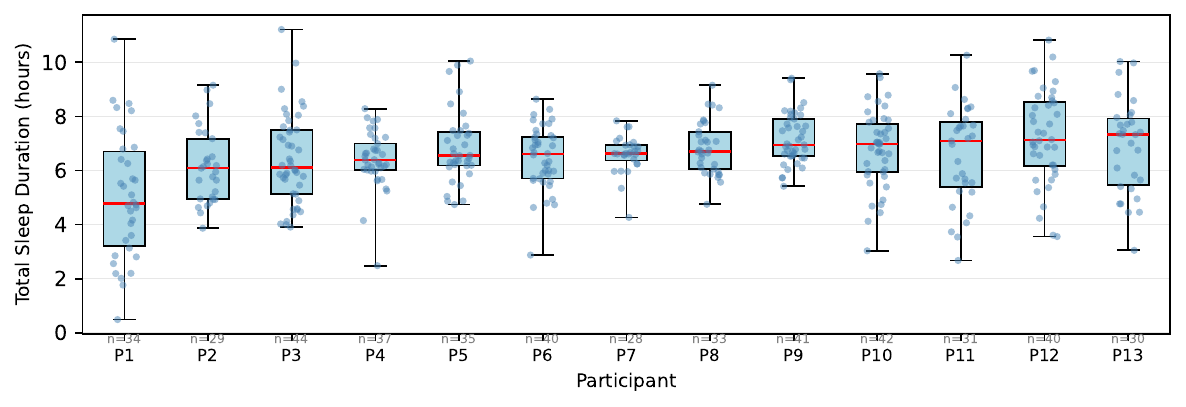}\\[1pt]
  \includegraphics[width=\columnwidth]{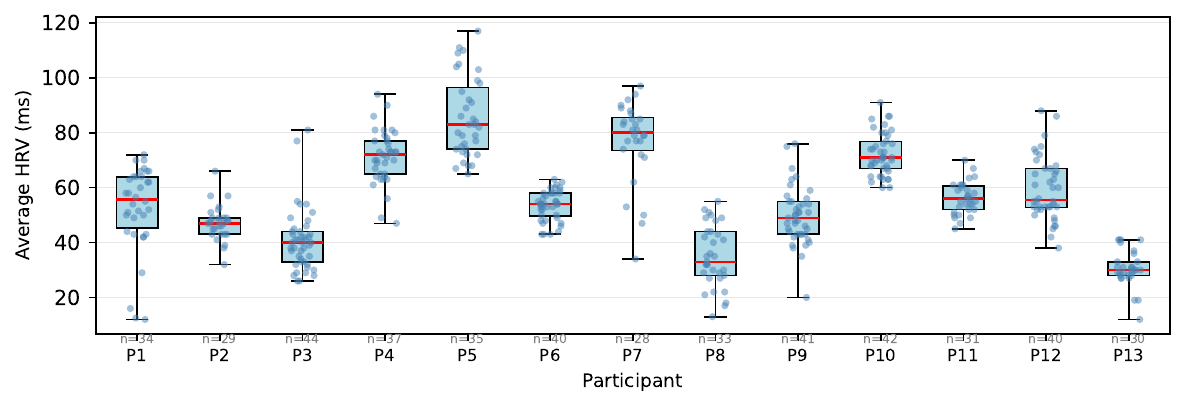}\\[1pt]
  \includegraphics[width=\columnwidth]{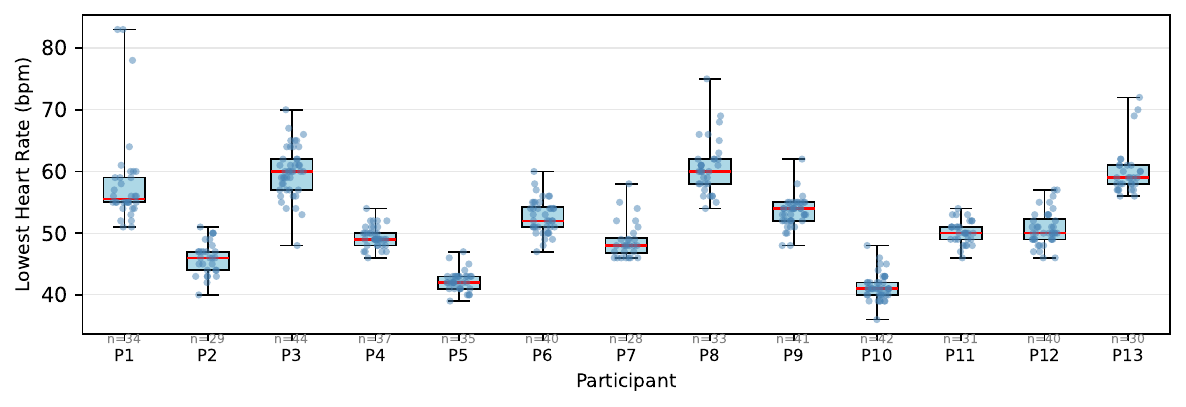}\\[1pt]
  \includegraphics[width=\columnwidth]{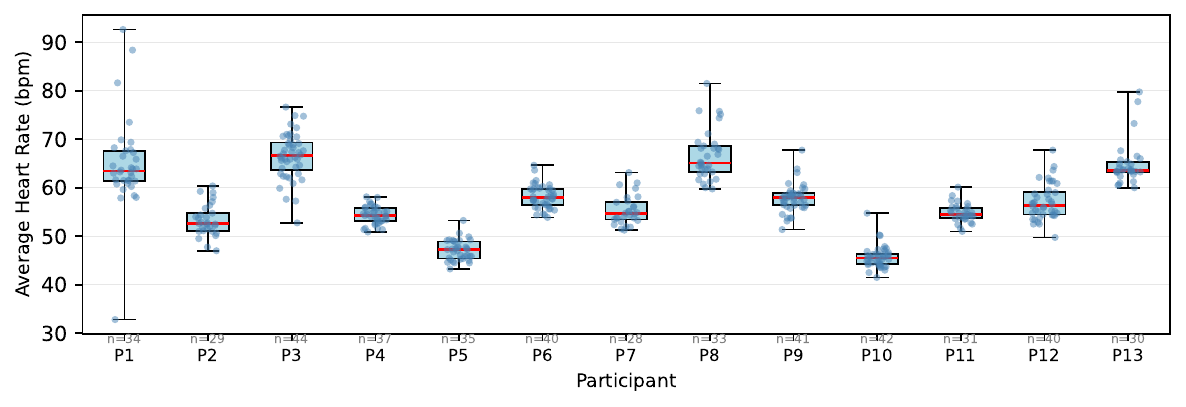}
  \caption{Distribution of target variables for each participant. Box plots show the median (red line), interquartile range (box), and minimum to maximum (whiskers).
  Blue dots represent individual daily data.
  Participants are ordered by ascending median of Total Sleep (P1--P13).}
  \label{fig:target_distribution}
\end{figure}

The target sleep indicators are:
total sleep duration (Total Sleep; mean 6.6~h, SD 1.5),
average heart rate variability (Avg HRV; mean 56.3~ms, SD 18.7),
lowest heart rate (Lowest HR; mean 51.8~bpm, SD 7.2),
and average heart rate (Avg HR; mean 57.4~bpm, SD 7.9).
Figure~\ref{fig:target_distribution} shows the distribution of each target variable per participant.
Participants are labeled P1--P13 in ascending order of their median Total Sleep.
Inter-participant variability is large; for example, the median Total Sleep ranges from 4.8 to 7.3~hours.
Intra-participant variability also differs across individuals, with some participants showing large day-to-day fluctuations.

\section{Experimental Results and Discussion}

\subsection{Evaluation Method}

Given the large inter-individual variability observed in sleep indicators (Fig.~\ref{fig:target_distribution}), a person-dependent modeling strategy was adopted: a classifier (Random Forest) was trained for each participant.
We adopted Leave-One-Day-Out cross-validation (LODOCV) for evaluation.
In LODOCV, one day's data is used as the test set and the remaining days' data are used for training.
That is, all 15 samples collected on the held-out day (three sessions $\times$ five tasks) are used exclusively for testing and never appear in the training set.
This procedure is repeated for all days to obtain predicted probabilities for each sample.
As evaluation metrics, we used PR-AUC (area under the Precision-Recall curve) and Recall@25\%, which are suitable for class imbalance.
PR-AUC summarizes how accurately the classifier can rank the positive (problematic) class,
and the baseline of a random classifier equals the positive class ratio (approximately 0.25).
Therefore, a PR-AUC $> 0.25$ indicates classification performance above chance.
Recall@25\% is the proportion of actual positives among the top 25\% of samples ranked by predicted probability,
indicating ``how many of the actual problematic cases are captured when selecting the top 25\% at highest risk'' from a clinical perspective.
The baseline for a random classifier is 0.25.

\subsection{Experimental Results}

% 予測対象の4変数はDatasetセクションで定義済み
The handwriting tasks consist of five types: circle, triangle, square, a positive phrase, and a negative phrase.
The sessions were conducted three times per day: after waking, after lunch, and before bedtime.

Table~\ref{tab:overall_results} shows the PR-AUC and Recall@25\% (mean across all tasks, user mean $\pm$ standard deviation)
and the results of the Wilcoxon signed-rank test for each target variable.
The Wilcoxon test evaluated whether each user's mean value significantly exceeded the baseline (0.25) using a one-sided test.
FDR correction (Benjamini-Hochberg method, $q = 0.05$) was applied to the 4 target variables $\times$ 2 metrics = 8 comparisons,
and all 8 tests remained significant.
This indicates that, under both metrics, sleep-related cardiac indicators and total sleep duration
can be identified from handwriting features with accuracy above random chance.
Lowest HR showed the highest classification performance (PR-AUC 0.438),
followed by Avg HRV (0.391), Avg HR (0.365), and Total Sleep (0.352).

\begin{figure}[p]
  \centering
  \begin{subfigure}[b]{0.68\textwidth}
    \centering
    \includegraphics[width=\textwidth]{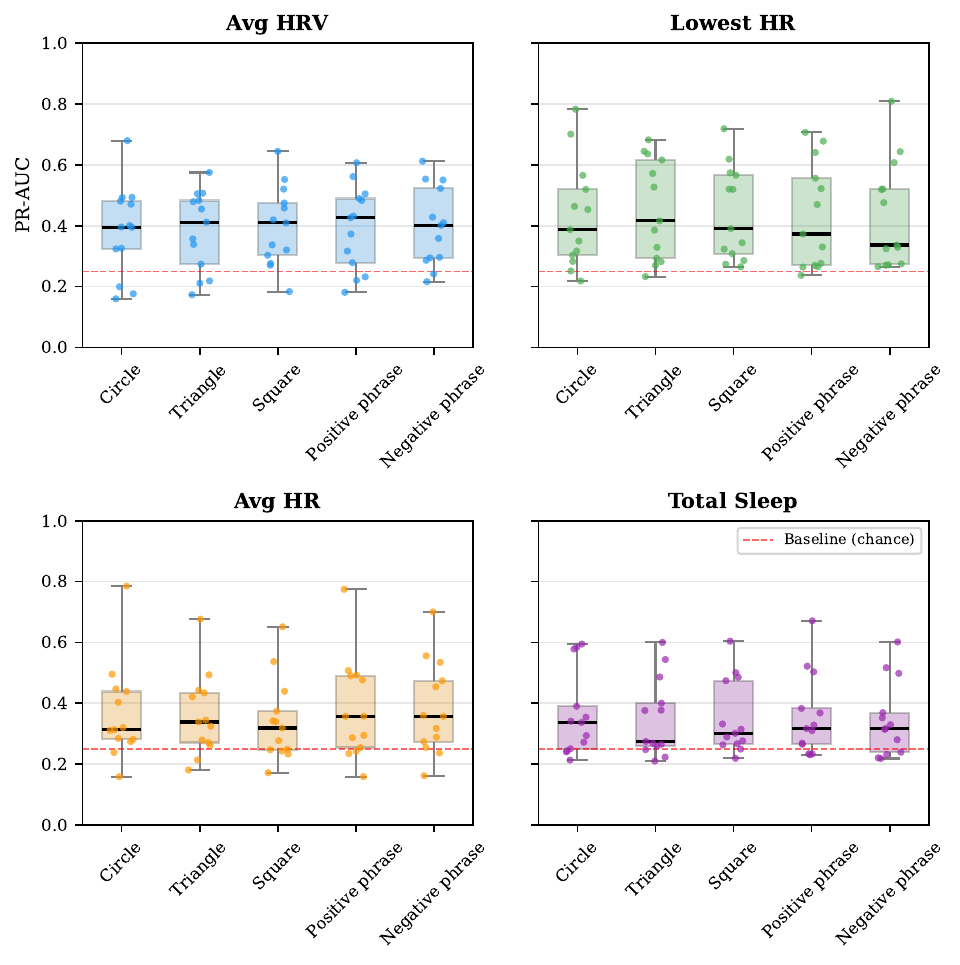}
    \caption{PR-AUC}
    \label{fig:prauc_task}
  \end{subfigure}
  \\[2pt]
  \begin{subfigure}[b]{0.68\textwidth}
    \centering
    \includegraphics[width=\textwidth]{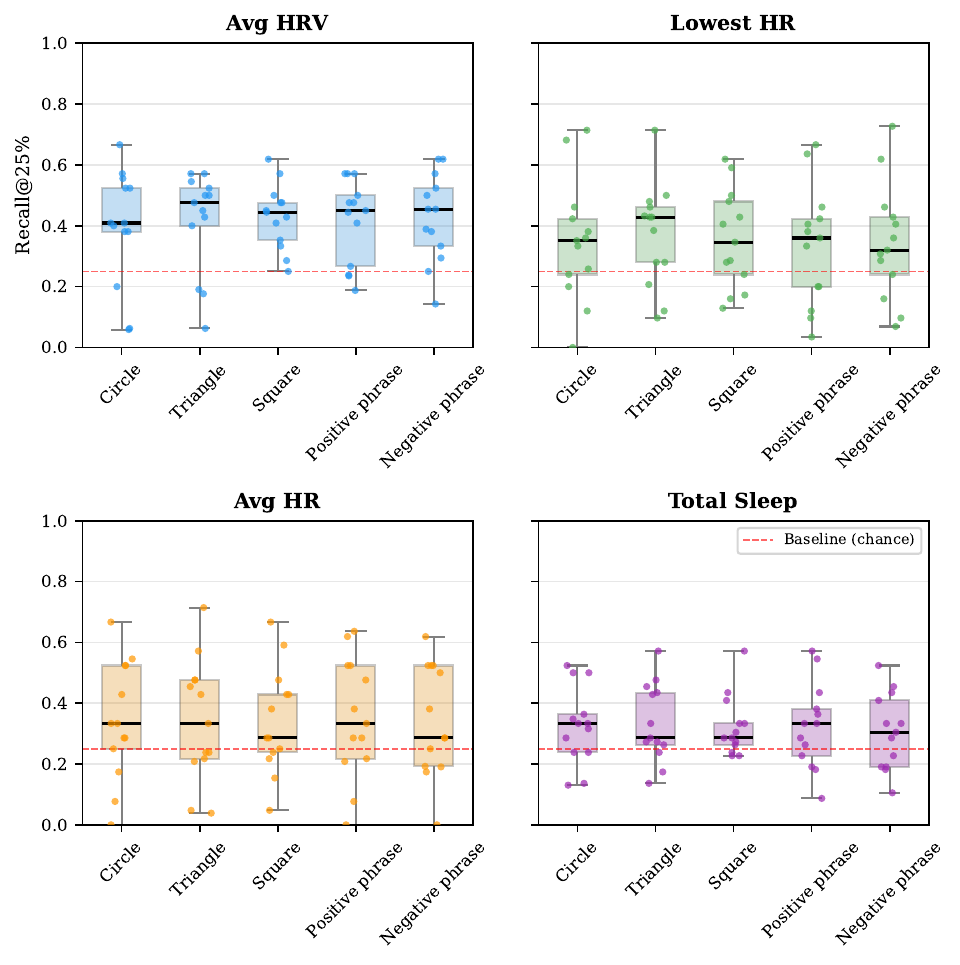}
    \caption{Recall@25\%}
    \label{fig:recall25_task}
  \end{subfigure}
  \caption{Classification performance by task (each target variable). Each point represents one user's result. The red dashed line indicates the baseline (0.25).}
  \label{fig:task_results}
\end{figure}

Figure~\ref{fig:task_results}~(a) and (b) show the distribution of PR-AUC and Recall@25\% by task.
Each point represents one user's value, and box plots show the distribution across users.
The red dashed line represents the baseline (0.25).
To assess whether there were differences in classification performance across tasks for each target variable, Friedman tests were conducted (Table~\ref{tab:friedman}).
After FDR correction (Benjamini-Hochberg method, 16 comparisons across 4 target variables $\times$ 2 metrics $\times$ 2 effects),
no significant differences were observed for any combination of target variable and metric.
This suggests that equivalent classification is achievable from any handwriting task.

% Friedman検定の結果（表~\ref{tab:friedman}），
% タスク間で両指標ともに大きな差は見られなかった．
% 他の7条件（3ターゲット$\times$2指標 + Avg HRのRecall@25\%）では
% タスクの主効果は有意でなかった（$p > 0.05$）．

\begin{figure}[p]
  \centering
  \begin{subfigure}[b]{0.68\textwidth}
    \centering
    \includegraphics[width=\textwidth]{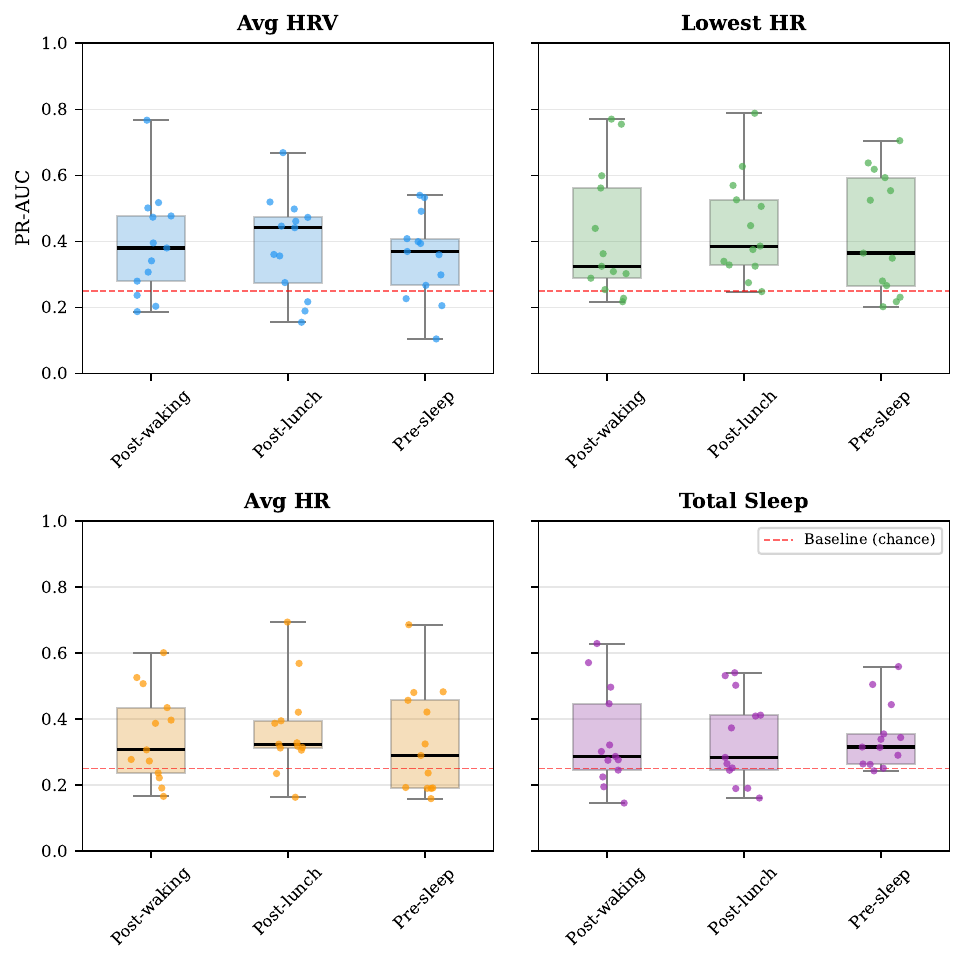}
    \caption{PR-AUC}
    \label{fig:prauc_timing}
  \end{subfigure}
  \\[2pt]
  \begin{subfigure}[b]{0.68\textwidth}
    \centering
    \includegraphics[width=\textwidth]{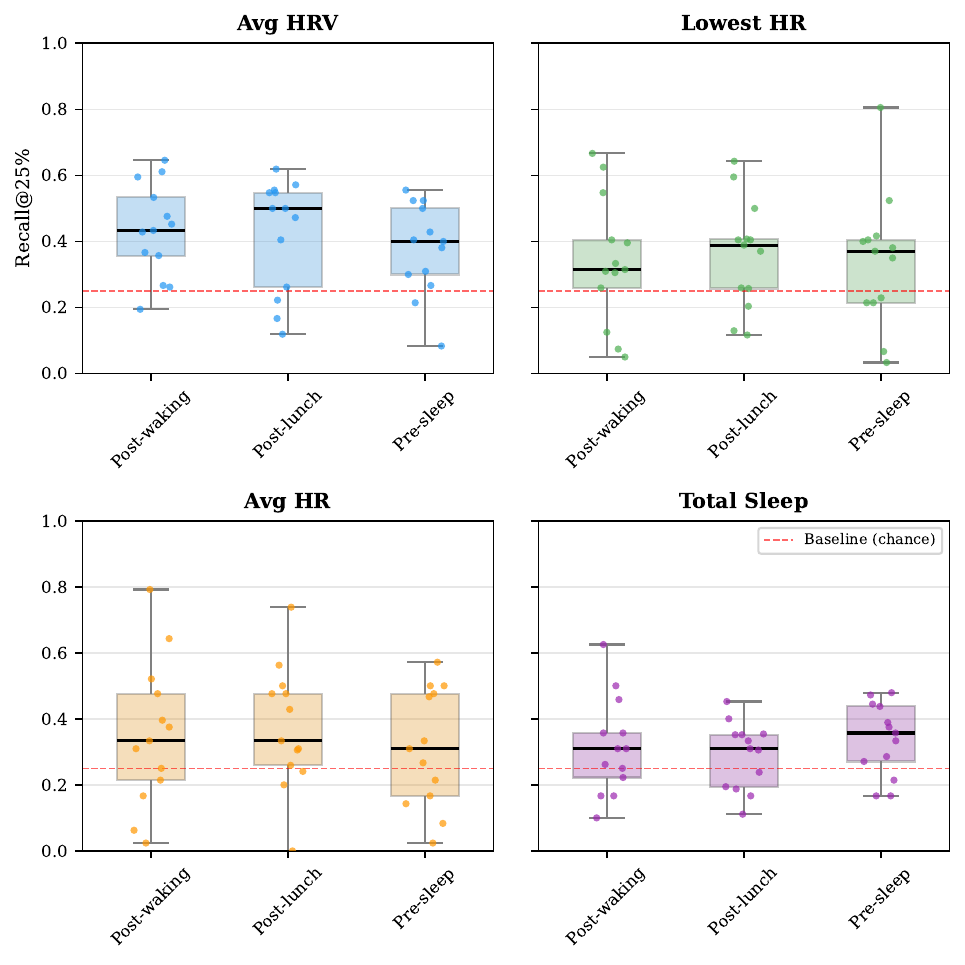}
    \caption{Recall@25\%}
    \label{fig:recall25_timing}
  \end{subfigure}
  \caption{Classification performance by timing (each target variable). Each point represents one user's result. The red dashed line indicates the baseline (0.25).}
  \label{fig:timing_results}
\end{figure}

Figure~\ref{fig:timing_results}~(a) and (b) show the distribution by timing.
Similarly, no significant timing effect was observed for any target variable after FDR correction (Table~\ref{tab:friedman}).
This suggests that comparable classification performance can be obtained
at any of the three timings: after waking, after lunch, and before bedtime.

\begin{table}[t]
  \centering
  \caption{PR-AUC and Recall@25\% for each target variable (user mean $\pm$ SD) with Wilcoxon test results.
  $p$: raw p-value; $q$: FDR-corrected p-value (Benjamini-Hochberg, 8 comparisons).
  The baseline (random classifier) is 0.25 for both metrics.}
  \label{tab:overall_results}
  \small
  \begin{tabular}{@{}lcccc@{\hspace{8pt}}cccc@{}}
    \toprule
     & \multicolumn{4}{c}{PR-AUC} & \multicolumn{4}{c}{Recall@25\%} \\
    \cmidrule(lr){2-5} \cmidrule(lr){6-9}
    Target variable & mean $\pm$ SD & $p$ & $q$ & sig. & mean $\pm$ SD & $p$ & $q$ & sig. \\
    \midrule
    Avg HRV     & $0.391 \pm 0.131$ & 0.0023 & 0.0046 & ** & $0.416 \pm 0.139$ & 0.0023 & 0.0046 & ** \\
    Lowest HR   & $0.438 \pm 0.164$ & 0.0001 & 0.0008 & *** & $0.350 \pm 0.174$ & 0.0327 & 0.0374 & * \\
    Avg HR      & $0.365 \pm 0.146$ & 0.0031 & 0.0050 & ** & $0.344 \pm 0.185$ & 0.0471 & 0.0471 & * \\
    Total Sleep & $0.352 \pm 0.125$ & 0.0023 & 0.0046 & ** & $0.322 \pm 0.108$ & 0.0199 & 0.0265 & * \\
    \bottomrule
  \end{tabular}
\end{table}

\begin{table}[t]
  \centering
  \caption{Results of Friedman tests. Tests were conducted for task effect (5 levels) and timing effect (3 levels) on PR-AUC and Recall@25\% respectively (16 tests in total). $q$: FDR-corrected p-value (Benjamini-Hochberg, 16 comparisons). None remained significant after correction.}
  \label{tab:friedman}
  \small
  \begin{tabular}{@{}lccc@{\hspace{6pt}}ccc@{\hspace{10pt}}ccc@{\hspace{6pt}}ccc@{}}
    \toprule
     & \multicolumn{6}{c}{Task effect} & \multicolumn{6}{c}{Timing effect} \\
    \cmidrule(lr){2-7} \cmidrule(lr){8-13}
     & \multicolumn{3}{c}{PR-AUC} & \multicolumn{3}{c}{R@25\%} & \multicolumn{3}{c}{PR-AUC} & \multicolumn{3}{c}{R@25\%} \\
    \cmidrule(lr){2-4} \cmidrule(lr){5-7} \cmidrule(lr){8-10} \cmidrule(lr){11-13}
    Target & $\chi^2$ & $p$ & $q$ & $\chi^2$ & $p$ & $q$ & $\chi^2$ & $p$ & $q$ & $\chi^2$ & $p$ & $q$ \\
    \midrule
    Avg HRV     & 2.40 & 0.663 & 0.847 & 2.26 & 0.688 & 0.847 & 0.46 & 0.794 & 0.907 & 6.84 & 0.033 & 0.344 \\
    Lowest HR   & 6.09 & 0.192 & 0.680 & 3.31 & 0.507 & 0.847 & 4.31 & 0.116 & 0.619 & 1.10 & 0.576 & 0.847 \\
    Avg HR      & 9.85 & 0.043 & 0.344 & 0.40 & 0.982 & 0.982 & 2.92 & 0.232 & 0.680 & 0.91 & 0.633 & 0.847 \\
    Total Sleep & 2.65 & 0.619 & 0.847 & 0.82 & 0.935 & 0.982 & 1.08 & 0.584 & 0.847 & 2.73 & 0.255 & 0.680 \\
    \bottomrule
  \end{tabular}
\end{table}

Additionally, large individual differences in PR-AUC across users can be observed in Fig.~\ref{fig:task_results}~(a) and Fig.~\ref{fig:timing_results}~(a).
For example, for Lowest HR, one user achieved a notably high PR-AUC of 0.729,
while several users remained near the baseline.
The pattern of individual differences varies by target variable;
a user who shows high performance for Lowest HR does not necessarily show high performance for Total Sleep.

\subsection{Discussion}

% Paragraph 1: Why can handwriting predict sleep?
The finding that handwriting features achieved significant classification performance
for all four sleep variables is consistent with the hypothesis that
changes in sleep quality affect daytime fine motor control via the autonomic nervous system.
The parameters of the Sigma-Lognormal model (reaction time $t_0$, lognormal parameter $D$, etc.)
reflect neuromuscular control characteristics~\cite{plamondon2021lognormality},
and changes in recovery state due to sleep may be reflected in these parameters.
Unlike prior work that relied on basic kinematics such as writing speed and fluency~\cite{jasper2009circadian},
the Sigma-Lognormal model may capture subtler motor variations that reflect daily recovery fluctuations.

% Paragraph 2: Why cardiac variables > Total Sleep?
Cardiac-related variables (Lowest HR: PR-AUC 0.438, Avg HRV: 0.391, Avg HR: 0.365)
showed higher classification performance than Total Sleep (0.352).
Heart rate and HRV are indicators that directly reflect the recovery state of the autonomic nervous system,
and fine motor control has been reported to be associated with autonomic regulation, with poor sleep increasing sympathetic tone and impairing fine motor skills~\cite{venevtseva2024sleep}.
In contrast, total sleep duration is an indicator of sleep ``quantity,''
and its correspondence with autonomic recovery is more indirect.
For example, long sleep with poor quality results in insufficient recovery,
while short but deep sleep can lead to high recovery.
This difference in directness is considered to be reflected in the classification performance gap.

% Paragraph 3: Task/timing independence
The absence of significant differences in classification performance across tasks and timings after FDR correction
suggests that the influence of sleep is not limited to specific motor patterns (e.g., drawing circles)
but permeates handwriting motor control in general.
Since the Sigma-Lognormal model captures dynamic properties of movement rather than character shape,
it can be interpreted as being independent of task type.
From a practical standpoint, this means that users do not need to perform specific tasks,
and prediction from everyday handwriting activities is feasible.

% Paragraph 4: Positioning relative to prior work
Existing handwriting analysis studies (ADHD classification~\cite{ADHD}, neurodegenerative disease diagnosis~\cite{pirlo2015neurodegeneration}, etc.)
primarily capture between-group differences between clinical and healthy populations.
In contrast, this study is fundamentally different in that it captures day-to-day variations within healthy individuals.
Between-group differences tend to have relatively large effect sizes,
whereas daily within-person variations are subtle and more difficult to detect.
The fact that this study was able to detect such subtle variations with statistical significance
suggests the richness of information contained in handwriting features.
It should be noted that detecting such small physiological variations under natural living conditions
represents a fundamentally more challenging problem with inherently smaller effect sizes
than between-group disease classification.

% Paragraph 5: Individual differences — issues for future investigation
On the other hand, the large individual differences in classification performance across users
represent an issue requiring further investigation.
Possible causes include individual characteristics of handwriting (pen pressure, writing speed, etc.),
individual differences in the magnitude of sleep variation,
and differences in motivation for experiment participation.
Additionally, the participants in this study were university students,
whose irregular lifestyle patterns (e.g., variable class schedules, part-time work)
may have acted as confounding factors.
Nevertheless, some users achieved high classification performance
(e.g., PR-AUC of 0.729 for Lowest HR),
demonstrating the potential of within-user models.
Future work should expand the number and diversity of participants (age groups, occupations, etc.)
and deepen the understanding of which user characteristics are associated with prediction performance.

\section{Conclusion}

Using Sigma-Lognormal model-based handwriting features,
we evaluated a binary classification framework for detecting low-recovery days
with 28-day in-the-wild data from 13 participants.
LODOCV evaluation confirmed that PR-AUC significantly exceeded the baseline (0.25)
for all four sleep variables (highest: Lowest HR 0.438),
with all tests remaining significant after FDR correction.
Furthermore, no significant differences in classification performance were observed
across task types or session timings after FDR correction,
suggesting that the proposed approach is largely robust to task and timing variations within the evaluated conditions.

These results constitute the first empirical evidence that handwriting features
can be used not only for disease classification but also for estimating
daily health fluctuations in healthy individuals,
suggesting the feasibility of non-invasive sleep monitoring using tablet devices.

Future work includes verifying generalizability through expanding the number and diversity of participants,
analyzing factors contributing to individual differences,
and validating applicability to natural handwriting activities (e.g., note-taking) in educational settings.

% === Acknowledgements / Disclosure（arXiv版のみ表示） ===
\ifarxiv
\begin{credits}
\subsubsection{\ackname}
The authors would like to express their sincere gratitude to Mr. Toshihiko Horie and Mr. Takahiro Yamamoto of Wacom Co., Ltd. for lending the Wacom IoT Paper used in this study. This research was supported by the JST research program CRONOS (Grant No. JPMJCS24K4).
\subsubsection{\discintname}
The authors have no competing interests to declare that are
relevant to the content of this article.
\end{credits}
\fi
%
% ---- Bibliography ----
%
% BibTeX users should specify bibliography style 'splncs04'.
% References will then be sorted and formatted in the correct style.
%
%\clearpage
\bibliographystyle{splncs04}
\bibliography{mybibliography}
\end{document}